\documentclass[aps,prd,preprint,floatfix,12pt,showpacs,superscriptaddress,tightenlines,nofootinbib,longbibliography,a4paper]{revtex4-1}
\usepackage{amsfonts}
\usepackage{amsmath}
\usepackage[markup=nocolor]{changes}
\usepackage{graphicx}
\usepackage{hyperref}
\usepackage{multirow}
\usepackage{url}
\usepackage{xspace}

\newcommand*\oline[1]{
  \hspace*{0.2em}
  \vbox{
    \kern-0.35ex
    \hrule height 0.4pt
    \kern0.35ex
    \hbox{
      \kern-0.5em
      \ifmmode#1\else\ensuremath{#1}\fi
      \kern-0.0em
}}}
\newcommand{\CP}{$ \mathcal{CP} $}
\newcommand{\HEPfit}{\texttt{HEPfit}\xspace}
\newcommand{\invcomm}[1]{``#1''}
\newcommand{\nn}{\nonumber}

\hypersetup{pdfstartview=FitV,colorlinks=true,linkcolor=blue,citecolor=red,filecolor=black,urlcolor=blue}

\begin{document}

\title{Next-to-leading order unitarity fits in Two-Higgs-Doublet models with soft $\mathbb{Z}_2$ breaking}

\author{Vincenzo Cacchio}
\email{vincenzo.cacchio@roma1.infn.it}

\author{Debtosh Chowdhury}
\email{debtosh.chowdhury@roma1.infn.it}

\author{Otto Eberhardt}
\email{otto.eberhardt@roma1.infn.it}

\affiliation{Istituto Nazionale di Fisica Nucleare,
Sezione di Roma, Piazzale Aldo Moro 2, I-00185 Roma, Italy}

\author{Christopher W. Murphy}
\email{christopher.murphy@sns.it}

\affiliation{Scuola Normale Superiore,
Piazza dei Cavalieri 7, I-56126 Pisa, Italy}

\date{\today}

\begin{abstract}
\noindent
We fit the next-to-leading order unitarity conditions to the Two-Higgs-Doublet model with a softly broken $\mathbb{Z}_2$ symmetry. In doing so, we alleviate the existing uncertainty on how to treat higher order corrections to quartic couplings of its Higgs potential. A simplified approach to implementing the next-to-leading order unitarity conditions is presented. These new bounds are then combined with all other relevant constraints, including the complete set of LHC Run I data. The upper $95\%$ bounds we find are $4.2$ on the absolute values of the quartic couplings, and $235$ GeV ($100$ GeV) for the mass degeneracies between the heavy Higgs particles in the type I (type II) scenario. In type II, we exclude an unbroken $\mathbb{Z}_2$ symmetry with a probability of $95\%$. All fits are performed using the open-source code \texttt{HEPfit}.
\end{abstract}

\maketitle

\section{Introduction}
\label{sec:intro}

Run I of the Large Hadron Collider (LHC) has concluded with the discovery of the last missing piece of the Standard Model (SM) -- the Higgs boson \cite{Aad:2012tfa,Chatrchyan:2012xdj}. It has tested the validity of the SM in a previously unexplored regime of energy, and has not found any significant deviations from the SM, hinting at a gap in the mass spectrum from the SM to whatever lies beyond it. This picture is consistent with the absence of indications of New Physics (NP) coming from indirect searches (e.g. electroweak precision or flavour observables). Certainly Run I of the LHC did not address all the shortcomings of the SM -- among which are the hierarchy problem, dark matter and an explanation for the flavour pattern. In order to find a solution for these problems the presence of NP is inevitable. One of the key questions for Run II of the LHC is then at what scale the NP appears. The measured value of the Higgs mass, $m_h \approx 125$ GeV \cite{Aad:2015zhl}, hints at an answer to this: it tells us that the Higgs potential of the SM is not stable up to very high energy scales~\cite{Degrassi:2012ry,Buttazzo:2013uya}. Thus one has to introduce an additional mechanism or new degrees of freedom to stabilize the Higgs potential if one wants to exclude the possibility of vacuum metastability.
Furthermore, to keep the SM Higgs mass \textit{natural}, new degrees of freedom around the TeV scale are required \cite{Barbieri:1987fn}.

One well motivated direction for discovering physics beyond the SM is to search for additional Higgs bosons. These particles often arise in natural theories of electroweak symmetry breaking, e.g. the Higgs sector of the minimal supersymmetric Standard Model (MSSM)~\cite{Drees:2004jm,Baer:2006rs,Martin:1997ns}, twin Higgs models~\cite{Burdman:2014zta,Burdman:2008ek,Burdman:2006tz,Chacko:2005un,Chacko:2005pe}, composite Higgs models~\cite{Panico:2015jxa,Agashe:2004rs}. Also, there is no fundamental reason for the minimality of the SM scalar sector, and multiple generations are known to exist in the fermion sector. Furthermore, the uncertainties in the SM Higgs coupling measurement \cite{Khachatryan:2016vau} do not exclude the presence of additional scalars.

After the SM, the simplest and most straightforward extension of the SM is the addition of another Higgs doublet, the so-called Two-Higgs-Doublet model (2HDM) \cite{Lee:1973iz,Gunion:2002zf,Branco:2011iw}, which has been analysed in great detail in the literature, see for instance \cite{Chen:2013kt,Chiang:2013ixa,Grinstein:2013npa,Barroso:2013zxa,Coleppa:2013dya,Eberhardt:2013uba,Belanger:2013xza,Chang:2013ona,Cheung:2013rva,Celis:2013ixa,Wang:2013sha,Baglio:2014nea,Inoue:2014nva,Dumont:2014wha,Kanemura:2014bqa,Ferreira:2014sld,Broggio:2014mna,Dumont:2014kna,Bernon:2014nxa,Chen:2015gaa,Chowdhury:2015yja,Craig:2015jba,Bernon:2015qea,Bernon:2015wef}. It is interesting to study the unitarity bounds in the 2HDM because the scale at which new particles are expected to appear based on naturalness arguments is the same scale as the Lee, Quigg, Thacker upper limit~\cite{Lee:1977yc, Lee:1977eg} on the Higgs mass in the SM, which is of order of $1$ TeV. In fact, there exists a large number of works studying the tree-level unitarity bounds on the quartic couplings, $\lambda_i$, and Higgs masses of the 2HDM, see e.g. \cite{Casalbuoni:1986hy,Maalampi:1991fb,Kanemura:1993hm,Akeroyd:2000wc,Ginzburg:2005dt,Horejsi:2005da,Haber:2010bw}. Unlike in the SM, extracting the bounds on the masses of the Higgs boson from the bounds on the quartic couplings is not straightforward because in the 2HDM the quartic couplings are in general functions of more parameters than just the masses of the Higgs bosons and their corresponding vacuum expectation values (VEVs). Recently, the perturbative unitarity bounds in the $\mathcal{CP}$-conserving softly-broken $\mathbb{Z}_2$ symmetric 2HDM were analyzed at the one-loop level~\cite{Grinstein:2015rtl}. This calculation settled a particular issue regarding how to estimate higher-order effects on available upper limits on the quartic couplings: In the SM, the unitarity bounds had been determined beyond the leading order (LO), and the typical result was that the bounds on the (RG-improved) quartic coupling of the SM were improved by a factor of a few with respect to the tree-level analysis \cite{Durand:1992wb,Maher:1993vj,Durand:1993vn,Nierste:1995zx}. While it was known that the tree-level unitarity bounds in NP models were likely to be overly conservative, there was no well defined way to implement stricter bounds using only tree-level results. Inspired by the SM results, it was advocated to re-scale the tree-level conditions by a factor of $1/4$ to estimate higher order contributions \cite{Baglio:2014nea}. A renormalization group analysis at next-to-leading order (NLO) confirmed this prescription if one wants a stable Higgs potential beyond $10$ TeV \cite{Chowdhury:2015yja}. However, now that an explicit NLO computation is at hand, this uncertainty on how to treat higher order corrections to the partial-wave amplitudes has been removed.

In this article, we improve on the results currently available in the literature in two main ways: regarding the unitary constraints, we go beyond the leading order precision by employing one-loop corrections which are enhanced, $
\mathcal{O}(\lambda_i\lambda_j/16\pi^2)$, in the limit $s \gg |\lambda_i|v^2 \gg M_W^2$, $s \gg m_{12}^2$ to all the $2 \to 2$ longitudinal vector boson and Higgs boson scattering amplitudes. Secondly, we perform global parameter fits including the most up-to-date Run I ATLAS and CMS results, rather than only using a handful of benchmark scenarios, which might not cover the whole spectrum of interesting features.
 
The structure of this article is as follows:
We give a short introduction to the model and its constraints in sections \ref{sec:model} and \ref{sec:constraints}, respectively. The statistical framework is presented in section \ref{sec:hepfit}. Section \ref{sec:results} contains the results of the fits. We conclude in section \ref{sec:conclusions}. Supplementary figures can be found in \hyperref[sec:appendixA]{Appendix A}, while we list the formulae for the NLO unitarity criteria and the fit inputs in \hyperref[sec:appendixB]{Appendix B} and \hyperref[sec:appendixC]{Appendix C}, respectively.

\section{Model}
\label{sec:model}

The two-Higgs-doublet model (2HDM)~\cite{Lee:1973iz,Gunion:2002zf,Branco:2011iw} is a simple and straightforward extension of the Standard Model (SM), obtained through the addition of a second Higgs doublet to the SM field content. A characteristic of general 2HDMs is the existence of flavour-changing neutral currents (FCNC) mediated by tree-level exchange of neutral Higgs bosons. A natural way to eliminate these potentially dangerous FCNC is to require that the Yukawa interactions respect a discrete $\mathbb{Z}_2$ symmetry, which can be broken softly. The $\mathbb{Z}_2$ symmetry can be chosen in four independent ways, depending on the $\mathbb{Z}_2$ charge assignments for quarks and charged leptons; this lead to four different types of 2HDM which are referred to as type I, type II, type X (lepton specific) and type Y (flipped). The type II model is of particular interest because the Higgs sector of the MSSM is a 2HDM of type II. The models we focus on in this paper are the \CP-conserving 2HDM of type I and II with a softly broken $\mathbb{Z}_2$ symmetry.
The most general Higgs potential in a \CP-conserving 2HDM with a softly broken $\mathbb{Z}_2$ symmetry reads
\begin{align}
 V
 &=m_{11}^2\Phi_1^\dagger\Phi_1^{\phantom{\dagger}}
   +m_{22}^2\Phi_2^\dagger\Phi_2^{\phantom{\dagger}}
   -m_{12}^2 ( \Phi_1^\dagger\Phi_2^{\phantom{\dagger}}
              +\Phi_2^\dagger\Phi_1^{\phantom{\dagger}})
   +\tfrac12 \lambda_1(\Phi_1^\dagger\Phi_1^{\phantom{\dagger}})^2
   +\tfrac12 \lambda_2(\Phi_2^\dagger\Phi_2^{\phantom{\dagger}})^2
 \nonumber \\
 &\phantom{{}={}}
  +\lambda_3(\Phi_1^\dagger\Phi_1^{\phantom{\dagger}})
            (\Phi_2^\dagger\Phi_2^{\phantom{\dagger}})
  +\lambda_4(\Phi_1^\dagger\Phi_2^{\phantom{\dagger}})
            (\Phi_2^\dagger\Phi_1^{\phantom{\dagger}})
  +\tfrac12 \lambda_5 \left[ (\Phi_1^\dagger\Phi_2^{\phantom{\dagger}})^2
                      +(\Phi_2^\dagger\Phi_1^{\phantom{\dagger}})^2 \right],\label{eq:pot}
\end{align}
where $\Phi_1$ and $\Phi_2$ are the two complex $SU(2)_L$ Higgs doublets with hypercharge $Y = 1 / 2$ and the eight scalar potential parameters are real to avoid explicit  \CP-violation, with $ m_{12}^2 $ being the $\mathbb{Z}_2$ soft-breaking parameter. At the global minimum of the scalar potential $ V $ the neutral components of $\Phi_1$ and $\Phi_2$ acquire VEVs, $ v_1/\sqrt{2} $ and $ v_2/\sqrt{2} $, respectively, which are fixed by the minimization of the scalar potential and must satisfy $ v_1^2+v_2^2 \equiv v^2 \approx(246\,\text{GeV})^2$. The ratio of the two VEVs is defined as $ \tan\beta=v_2/v_1 $, where $ 0\leq\beta\leq\pi/2 $. Assuming no \CP-violation in the Higgs sector, the physical scalar spectrum consists of two \CP-even states $ h $ and $ H $ with $ m_{h}<m_{H} $, the \CP-odd state $ A $ and the charged state $ H^\pm $. The masses of these scalar bosons are denoted as $ m_\phi $ with $ \phi \in \{h,\,H,\,A,\, H^\pm \}$. Throughout this paper we interpret the observed Higgs resonance as the light \CP-even scalar $ h $ and thus treat $ m_h $ as fixed by measurements, $ m_h=125.09 \,\text{GeV} $ \cite{Aad:2015zhl}. We choose the independent  physical parameters of the model to be 
\begin{equation}
\tan\beta,\quad\beta-\alpha,\quad m_{12}^2,\quad m_{H},\quad m_{A},\quad m_{H^\pm},
\label{eq:SetOfParameters}
\end{equation}
where $ \alpha $ is the mixing angle of the neutral \CP-even 2HDM Higgs bosons. In this parametrization, the tree-level couplings of the Higgs bosons to vector bosons and fermions only depend on $ \tan\beta $ and $ \beta-\alpha $. Moreover, for $ \beta-\alpha=\pi/2 $ the couplings of $ h $ to SM fermions and vector bosons are SM-like and $ H $ does not couple to vector bosons at tree-level; the literature refers to this as the \textit{alignment limit}~\cite{Gunion:2002zf,Delgado:2013zfa,Craig:2013hca,Carena:2013ooa}.

Considering only the third generation of fermions, the Yukawa Lagrangian under the above-mentioned $\mathbb{Z}_2$ symmetry takes the following shape:
\begin{align}
 {\cal L}_Y =& -Y_t \oline Q_{\textit{\tiny{L}}} i\sigma_2 \Phi_2^* t_{\textit{\tiny{R}}} -Y_{b} \oline Q_{\textit{\tiny{L}}} \Phi_k b_{\textit{\tiny{R}}} -Y_{\tau} \oline L_{\textit{\tiny{L}}} \Phi_k \tau_{\textit{\tiny{R}}} + {\text{h.c.}},
  \label{eq:yuk}
\end{align}
where the top quark couples to $ \Phi_2 $ by convention and the index $k$ is $2$ in type I and $1$ in type II. The top Yukawa coupling is related to the SM value $ Y_t^\text{SM} $ by $ Y_t = Y_t^\text{SM}/\sin\beta$, while $ Y_f = Y_f^\text{SM}/\sin\beta$ in type I and $ Y_f = Y_f^\text{SM}/\cos\beta$ in type II for $f=b,\tau$.

\section{Constraints}
\label{sec:constraints}

In this section we list and discuss the theoretical and experimental constraints we impose on the 2HDM parameter space.
Since we want to combine them in a Bayesian fit, we list the priors on the parameters from \eqref{eq:SetOfParameters}:

\begin{center}
\begin{table}[h]
  \begin{tabular}{| l | c | c | c | c |}
    \hline
    \textbf{Parameter } & $\tan\beta$ & $\beta-\alpha$& $m_{12}^2$ & $m_{H}, m_{A}, m_{H^\pm}$\\ 
    \hline
    \textbf{Range}& $[0.25;100]$ & $[0;\pi]$ & $[-5\cdot 10^4;7\cdot 10^5]$ GeV$^2$ & $[130;1100]$ GeV\\
    \hline
    \end{tabular}
     \caption{Priors on the 2HDM parameters.}
\end{table}
\end{center}

\subsection{Theoretical constraints}
\label{sec:theoconstraints}

On the theory side, constraints on the 2HDM come from the following requirements:
\begin{itemize}
\item the Higgs potential must be bounded from below \cite{Deshpande:1977rw} between $M_Z$ and $750$ GeV,
\item the minimum of the Higgs potential at $ 246 $ GeV should be the global minimum \cite{Barroso:2013awa},
\item the 2HDM quartic couplings $ \lambda_i $ ($ i=1,\,2,\,3,\,4,\,5 $) and the Yukawa couplings are assumed to be perturbative (i.e. smaller than $4\pi$ and $\sqrt{4\pi}$ in magnitude, respectively) at least up to $750$ GeV,
\item the $S$-matrix of $2\to 2$ scattering processes for Higgs bosons and longitudinal vector bosons should be unitary up to NLO, and its NLO eigenvalues should not exceed the LO eigenvalues in magnitude~\cite{Grinstein:2015rtl}.
\end{itemize}
Requiring positivity and perturbativity of the couplings to hold at least up to $750$ GeV is motivated by the fact that this scale is well above the electroweak symmetry breaking scale and we can safely use the NLO unitarity conditions. For the renormalization group running we use NLO renormalization group equations (RGE) \cite{Chowdhury:2015yja}.

The first three bullet points have already been used in the literature and will be referred to as ``stability up to $750$ GeV'' in the following. The fourth set of constraints has never been applied in a general 2HDM fit, which is why we want to explain the details of our approach in the following, referring to \cite{Grinstein:2015rtl}.

The unitarity of the $S$-matrix leads to constraints on the partial wave amplitudes of a theory,
\begin{equation}
|a_{\ell}^{2 \to 2} - \tfrac{1}{2} i|^2 + \sum_{n > 2} |a_{\ell}^{2 \to n}|^2 = \tfrac{1}{4} ,
\end{equation}
where $a_{\ell}^{2 \to n}$ are the eigenvalues of the matrix of the $\ell$-th, $2 \to n$ partial wave amplitudes, $\mathbf{a}_{\ell}$. Considering only $2 \to 2$ scattering (and dropping the superscript) this constraint becomes an inequality,
\begin{equation}
\label{eq:ine}
|a_{\ell} - \tfrac{1}{2} i| \leq \tfrac{1}{2} .
\end{equation}
The 2HDM one-loop corrections necessary to use this inequality were recently computed in Ref.~\cite{Grinstein:2015rtl}. Prior to this computation, the inequalities $|\text{Re}(a_0)| \leq \tfrac{1}{2}$ or $|a_0| \leq 1$ were used to constrain the tree-level partial wave amplitudes \cite{Casalbuoni:1986hy,Maalampi:1991fb,Kanemura:1993hm,Akeroyd:2000wc,Ginzburg:2005dt,Horejsi:2005da,Haber:2010bw}. Comparing with the discussion of higher order corrections in the SM, stronger bounds were estimated and used for the 2HDM \cite{Baglio:2014nea,Chowdhury:2015yja}, but this ansatz was controversial. Having at hand the calculated NLO unitarity conditions, we can determine the upper bound on the quartic couplings without any ambiguity of method.

The computation of Ref.~\cite{Grinstein:2015rtl} was performed in the high energy limit, $s \gg |\lambda_i| v^2 \gg M_W^2$, $s \gg m_{12}^2$, where the $SU(2)_L \times U(1)_Y$ symmetry is manifest. In this limit, $\mathbf{a}_0$ is block diagonal at leading order, with blocks of definite weak isospin ($\sigma$) and hypercharge ($Y$) ($\mathbf{a}_{\ell > 0} = \mathbf{0}$ at leading order in this limit). Furthermore, the $\mathbb{Z}_2$-even and -odd states do not mix at tree-level, leading to smaller blocks. Due to the manifest symmetry at high energies, the calculation can be simplified by computing the amplitudes in the $\mathbb{Z}_2$ basis using the non-physical Higgs fields, $w_j^{\pm}$, $n_j^{(*)}$~\cite{Ginzburg:2005dt},
\begin{equation}
\Phi_j =
\begin{pmatrix}
w_j^+ \\
n_j + v_j / \sqrt{2}
\end{pmatrix} , \quad
n_j = \frac{h_j + i z_j}{\sqrt{2}}, \quad (j = 1,2) .
\end{equation}
The elements of $\mathbf{a}_0$ are given by
\begin{equation}
(\mathbf{a}_0)_{i,f} = \frac{1}{16 \pi s} \int_{-s}^0 \! dt \, \mathcal{M}_{i \otimes f}(s,t) ,
\end{equation}
where, for example,
\begin{align}
\label{eq:Mex}
\mathcal{M}_{\tfrac{1}{\sqrt{2}} (\Phi_1^{\dagger} \Phi_1) \otimes \tfrac{1}{\sqrt{2}} (\Phi_2^{\dagger} \tau^3 \Phi_2)} = \frac{1}{2} &\left( \mathcal{M}_{w_1^+ w_1^- \to w_2^+ w_2^-} - \mathcal{M}_{w_1^+ w_1^- \to n_2 n_2^*} \right. \\
&\left. + \mathcal{M}_{n_1 n_1^* \to w_2^+ w_2^-} - \mathcal{M}_{n_1 n_1^* \to n_2 n_2^*} \right). \nn
\end{align}

In general, the block diagonal structure of $\mathbf{a}_0$ does not hold beyond tree-level. However, it turns out that in the high energy limit, this structure is only broken by diagrams that correct the wavefunctions of the external legs, not by 1PI diagrams. Ref.~\cite{Grinstein:2015rtl} showed that the external wavefunction corrections are numerically subdominant with respect to the 1PI diagrams in some special cases. We confirm this and find it to be generalizable for all 2HDM scenarios with a softly broken $\mathbb{Z}_2$ symmetry. Due to this relative numerical unimportance, we neglect the external wavefunction corrections throughout this work. In this approximation, the one-loop eigenvalues take the following form,\footnote{Ref.~\cite{Grinstein:2015rtl} used the differential operator $\mathcal{D}_{\rm GMU} = 16 \pi^2 \mu^2 (d/d\mu^2)$ in its definition of the beta functions. In this work we use the traditional definition of the beta function, $\beta_{\lambda_i} \equiv \mathcal{D} \lambda_i = \mu (d\lambda_i / d\mu)$.}
\begin{align}
32 \pi a^{\text{even}}_{00\pm} &= B_1+ B_2 \pm \sqrt{(B_1-B_2)^2+4 B_3^2},  \\ 
32 \pi a^{\text{odd}}_{00\pm} &= 2 B_4 \pm 2 B_6, \nn \\ 
32 \pi a^{\text{even}}_{01\pm} &= B_7 + B_8 \pm \sqrt{(B_7-B_8)^2+4 B_9^2}, \nn  \\ 
32 \pi a^{\text{odd}}_{01\pm} &= 2 B_{13} \pm 2 B_{15}, \nn \\ 
32 \pi a^{\text{odd}}_{10} &= 2 B_{19}, \nn \\ 
32 \pi a^{\text{even}}_{11\pm} &= B_{20}+B_{21}\pm \sqrt{(B_{20}-B_{21})^2+4 B_{22}^2}, \nn \\
32 \pi a^{\text{odd}}_{11} &= 2 B_{30} , \nn
\end{align}
with the eigenvalues labeled as follows, $a_{Y \sigma \pm}^{\mathbb{Z}_2}$, and dropping the index $\ell=0$. $B_N$ is the block-diagonal element, $(\mathbf{a}_0)_{i,f}$, from Eq.~(B.N) in~\cite{Grinstein:2015rtl}, which can also be found in \hyperref[sec:appendixB]{Appendix B}. In order to satisfy unitarity, the $a_{Y \sigma \pm}^{\mathbb{Z}_2}$ have to individually fulfill the condition \eqref{eq:ine}. Note that at LO, the eigenvalues are related to the ones defined in \cite{Ginzburg:2005dt} by $a^{\mathbb{Z}_2}_{Y\sigma\pm}=-32\pi^2 \Lambda ^{\mathbb{Z}_2}_{Y\sigma\mp}$ for $\lambda_5>0$. 

Another constraint is the requirement that higher order corrections to the partial wave amplitudes are suppressed. In particular, following \cite{Grinstein:2015rtl} we define,
\begin{equation}
R_1^{\prime} = \frac{\left|a_{Y \sigma \pm}^{\mathbb{Z}_2, \text{NLO}}\right|}{\left|a_{Y \sigma \pm}^{\mathbb{Z}_2, \text{LO}}\right|} ,
\end{equation}
where the (N)LO label denotes the pure (N)LO contribution. Similar criteria were used in the perturbative unitarity analysis of the SM in Ref.~\cite{Durand:1993vn}. Assuming that the power series is perturbatively stable, we want to require the NLO contribution to be smaller than the LO expression, hence $R_1^{\prime}<1$. However, we need to avoid the exclusion of accidentally small leading-order contributions. (For instance, $a^{\text{odd},\text{LO}}_{10}=(\lambda_4-\lambda_3)/(8\pi )$ is small if $\lambda_3\approx \lambda_4$, while $a^{\text{odd},\text{NLO}}_{10}$ also depends on the other quartic couplings.) Therefore, we decided to use the $R_1^{\prime}$ criterion only if $|a^{\mathbb{Z}_2,\text{LO}}_{Y\sigma\pm}|>0.02\approx 1/(16\pi)$.

The 2HDM \textit{is} unitary, so let us explain what we mean when we say unitarity constraints. Inequality \eqref{eq:ine} requires the couplings of a theory to be smaller than a certain value in magnitude, or else the theory will no longer appear to be unitary at the finite order of the perturbative expansion to which we are working. In this sense both the ``perturbativity bound,'' $R_1^{\prime}$, and the ``unitarity bound'' test the same thing, namely where the breakdown of perturbation theory occurs.

\subsection{Experimental constraints}
\label{sec:expconstraints}

The experimental constraints included in our analysis are:
\begin{itemize}
\item the Peskin-Takeuchi parameters $ S $, $ T $, and $ U $ \cite{Haber:1993wf},
\item the $ h $ signal strengths,
\item the non-observation of $H$ and $A$ at the LHC and
\item the $B_s$ meson mass difference $\Delta m_{B_s}$ \cite{Geng:1988bq,Deschamps:2009rh} and the branching ratio $\mathcal{B}(\bar{B}\to X_s\gamma)$ \cite{Misiak:2015xwa}.
\end{itemize}
As we saw in the previous section, the 2HDMs introduce new Higgs bosons which couple to the gauge bosons and which, thereby, can give contributions, through loops, to the gauge boson self-energies. Thus, the 2HDMs yield new contributions to $ S $, $ T $ and $ U $ that generally move them away from their SM values. For the 2HDM predictions of the Peskin-Takeuchi parameters in the \CP-conserving limit we make use of the formulae of \cite{Haber:1993wf}. As input values for the oblique parameters $ S $, $ T $ and $ U $ and their correlation coefficients we take the most recent results obtained in a fit to electroweak precision data with \HEPfit \cite{deBlas:2016ojx}, see Table \ref{tab:STU} in \hyperref[sec:appendixC]{Appendix C}.

In order to confront the 2HDM with the latest ATLAS and CMS Run I data on Higgs signal strengths, we compute in the narrow-width approximation for each final state $ f\in\{\gamma\gamma,\,ZZ,\,WW,\,bb,\,\tau\tau\} $  the signal strengths\footnote{For the sake of simplicity, we refrain from writing obvious antiparticle and charge attributions explicitly.}
\begin{align}
\mu_\text{ggF+tth}^{f}&=\sum_{i=\textup{ggF},\,\textup{tth}}
\frac{{\sigma}_i^\text{2HDM}}{{\sigma}_i^\text{SM}}\cdot\frac{\mathcal{B}^\text{2HDM}(h{\to}f)}{\mathcal{B}^\text{SM}(h{\to}f)},\label{eq:sigstr1}\\
\mu_\text{VBF+Vh}^{f}&=\sum_{i=\text{VBF},\,\text{Vh}}
\frac{{\sigma}_i^\text{2HDM}}{{\sigma}_i^\text{SM}}\cdot\frac{\mathcal{B}^\text{2HDM}(h{\to}f)}{\mathcal{B}^\text{SM}(h{\to}f)},\label{eq:sigstr2}
\end{align}
having grouped the Higgs production modes in just two effective modes, $ \text{ggF}+\text{tth} $ and $ \text{VBF}+\text{Vh} $, where \invcomm{ggF}, \invcomm{tth}, \invcomm{VBF} and \invcomm{Vh} stand for \invcomm{gluon fusion}, ``$ t\bar{t} $ associated production'', \invcomm{vector boson fusion} and \invcomm{Higgstrahlung}, respectively. The SM Higgs boson production cross sections are taken at a centre-of-mass energy of $8\,\text{TeV}$ from \cite{Heinemeyer:2013tqa}; the SM branching ratios were calculated with HDecay 6.10 \cite{Djouadi:1997yw}. In order to express the 2HDM cross sections and branching ratios in terms of the SM ones, we make use of the formulae of \cite{Gunion:1989we} for the loop induced decays of the neutral Higgs bosons.
Central values, errors (Gaussian approximation) and correlations for the signal strengths in~\eqref{eq:sigstr1} and~\eqref{eq:sigstr2} were obtained from Fig.~13 and Table 14 of \cite{Khachatryan:2016vau} and can be found in Table \ref{tab:signalstrengths} in \hyperref[sec:appendixC]{Appendix C}.

Direct $ H $ and $ A $ searches are taken into account as follows: 
given the $ X\to H/A \to Y $ process, we define the ratio
\begin{align}
 R^{(X\to H/A \to Y)}_\text{\tiny Gauss} &= \frac{\sigma \mathcal{B}|_{\text{\tiny{theo}}} - (\sigma \mathcal{B}|_{\text{\tiny{95\%,obs}}}-\sigma \mathcal{B}|_{\text{\tiny{95\%,exp}}})}{\sigma \mathcal{B}|_{\text{\tiny{95\%,exp}}}}, \nonumber
\end{align}
where $\sigma \mathcal{B} =\sigma(X\to H/A)\cdot \mathcal{B}(H/A \to Y)$ and the subscripts denote the theoretical 2HDM value of $\sigma \mathcal{B}$ and its observed and expected exclusion limit at $ 95\% $ CL by the experiments. With this definition, we can assume the $R^{(X\to H/A \to Y)}_\text{\tiny Gauss}$ ratios to be Gaussian with a standard deviation of $1$. Note that these quantities depend on $m_{H/A}$; furthermore we neglect the error on $\sigma \mathcal{B}|_{\text{\tiny{95\%,exp}}}$.\\
The $H$ and $A$ search exclusion limits included in our analysis and their mass ranges, along with the exclusion plots from which they were digitalized, are listed in Table~\ref{tab:heavysearches} in \hyperref[sec:appendixC]{Appendix C}. Most SM Higgs production cross sections are taken from the LHC Higgs Cross Section Working Group \cite{LHCHXSWG}; the remaining ones are calculated with \textsc{HIGLU} $ 4.34 $ \cite{Spira:1995mt}, Sushi 1.5 \cite{Harlander:2012pb}, and Madgraph5 2.2.2 \cite{Alwall:2014hca}. The branching ratios were calculated with HDecay 6.10 \cite{Djouadi:1997yw} while the decay widths for both Higgs-to-Higgs decays and Higgs decays into a Higgs boson and a gauge boson are taken from \cite{Agashe:2014kda}.

From the plethora of flavour observables we only use the most relevant two for our 2HDM discussion: the mass difference in the $B_s$ meson system, $\Delta m_{B_s}$ and the branching fraction $\mathcal{B}(\bar{B}\to X_s\gamma)$. The former is calculated according to \cite{Geng:1988bq,Deschamps:2009rh} at LO. For the inclusive measurement of $\mathcal{B}(\bar{B}\to X_s\gamma)$, NNLO corrections are important \cite{Misiak:2015xwa}. As for fixed SM parameters this observable only depends on the two 2HDM parameters $\tan \beta$ and $m_{H^\pm}$, we store the $\mathcal{B}(\bar{B}\to X_s\gamma)$ values for various inputs of these two parameters in tables, and interpolate them linearly in the fits. A theoretical error of $7\%$ is applied, which corresponds to the uncertainty in the SM parameters.
The experimental inputs for the flavour observables can be found in Table \ref{tab:Flavour} in \hyperref[sec:appendixC]{Appendix C}.

\section{HEPfit}
\label{sec:hepfit}

As numerical set-up we use the open-source code \HEPfit \cite{hepfit}, interfaced with the release version of the Bayesian Analysis Toolkit (BAT) \cite{Caldwell:2009aa}. The former calculates all mentioned 2HDM observables and feeds them into the parallelized BAT, which applies the Bayesian fit with Markov chain Monte Carlo simulations. The complete global fit with all theoretical bounds runs for approximately $60$ hours with $12$ parallel chains generating $2\cdot 10^7$ iterations each. Adding the experimental observables as described above slows down the same fit to roughly $90$ hours.\\
A fundamental difference between the Bayesian and the frequentist approach is the treatment of fine-tuning: if one changes the parametrization of a model, flat priors on the former parameters usually do not translate into flat distributions of the new basis in a Bayesian fit. Some values for a new parameter might only be obtained by a very specific constellation of the old parameters, which in that sense would mean that they require a certain amount of fine-tuning. A frequentist fit is not sensitive to this bias, but one could argue that it is also less natural. \HEPfit makes use of the Bayesian approach assuming flat priors for the physical parameters \eqref{eq:SetOfParameters}, and the posterior distribution of the parameters in the Higgs potential \eqref{eq:pot} are ``distorted'' by the Jacobian of the change of parameters. However, the posterior intervals only have a well-defined meaning once experimental data is included, and that is when the dependence on the priors disappears. In the first part of the following section, when we only discuss theoretical constraints, the reader should bear in mind that our results depend on the priors (and thus on the parametrization). Also, we will present the $99.7$\% allowed regions for fits to only theoretical constraints, while after the inclusion of experimental data we show the $95.4$\% probability contours, which then have a statistical meaning.

\section{Results}
\label{sec:results}

\begin{figure}
   \begin{picture}(450,450)(0,0)
    \put(0,0){\includegraphics[width=450pt]{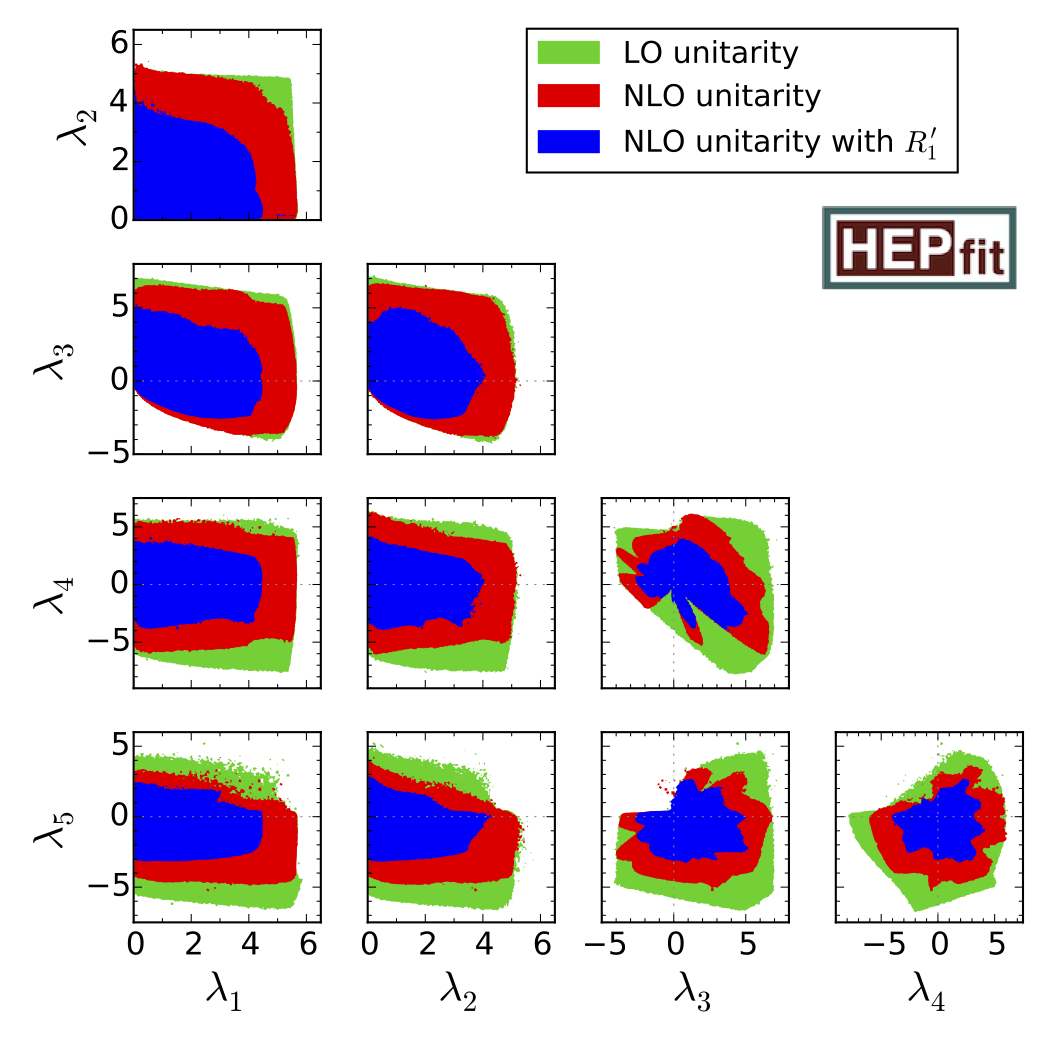}}
   \end{picture}
  \caption{$99.7\%$ probability regions in the $\lambda_i$ vs.~$\lambda_j$ planes. The green contours show the possible ranges if we impose a stable potential up to 750 GeV and unitarity at LO; the red contours mark the allowed regions if we take the NLO expressions instead, and the blue lines are obtained with the additional assumption that the ratio of NLO and LO unitarity is smaller than one (if the LO value is sufficiently large). The $\lambda_i$ values are at the scale $M_Z$.}
  \label{fig:lambdaswithunitarity}
\end{figure}

In the following we will present the results of our fits of theoretical and experimental constraints to the 2HDM of type I and II.
Before we address the physical 2HDM parameters we want to compare the effects of the unitarity constraints. As explained in Sec.~\ref{sec:theoconstraints}, we impose these bounds at a scale of $750$ GeV; thus a stable Higgs potential at least up to this scale is implicitly assumed. Nonetheless, all quantities shown in the figures are to be understood at the scale $M_Z$. In Fig.~\ref{fig:lambdaswithunitarity} we show the $99.7$\% probability regions for all $\lambda_i$ vs.~$\lambda_j$ planes with three different unitarity conditions: The green areas are allowed if we impose only LO unitarity, the red regions show the remaining parameter space if we use the NLO unitarity conditions, and the blue contours result from additionally requiring the NLO unitarity conditions to be perturbative ($R_1^{\prime}<1$). Generally one can see that perturbative NLO unitarity is a stronger constraint than NLO unitarity with arbitrary $R_1^{\prime}$, which itself is always stronger than LO unitarity. Numerically the largest possible absolute values for any of the quartic couplings from \eqref{eq:pot} are $8.10$, $7.21$, or $5.75$, if we apply LO, NLO or perturbative NLO unitarity, respectively.
Especially in the $\lambda_4$ vs.~$\lambda_3$ plane, but also in the $\lambda_5$ vs.~$\lambda_3$ and $\lambda_5$ vs.~$\lambda_4$ planes one can see that for particular constellations, NLO unitarity features sharp incisions towards the origin of the plane, whereas those indentations are absent if we use LO unitarity.

\begin{figure}
   \begin{picture}(450,370)(0,0)
   \put(0,0){\includegraphics[width=220pt]{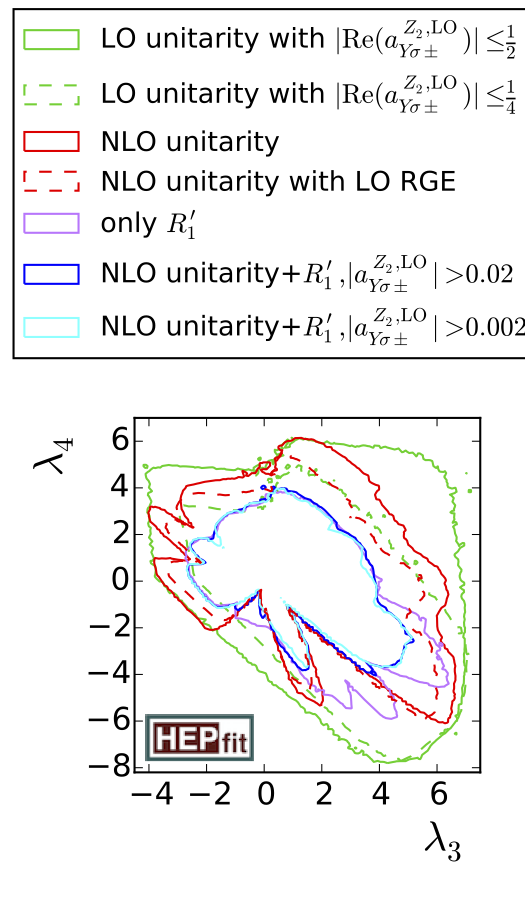}}
   \put(230,0){\includegraphics[width=220pt]{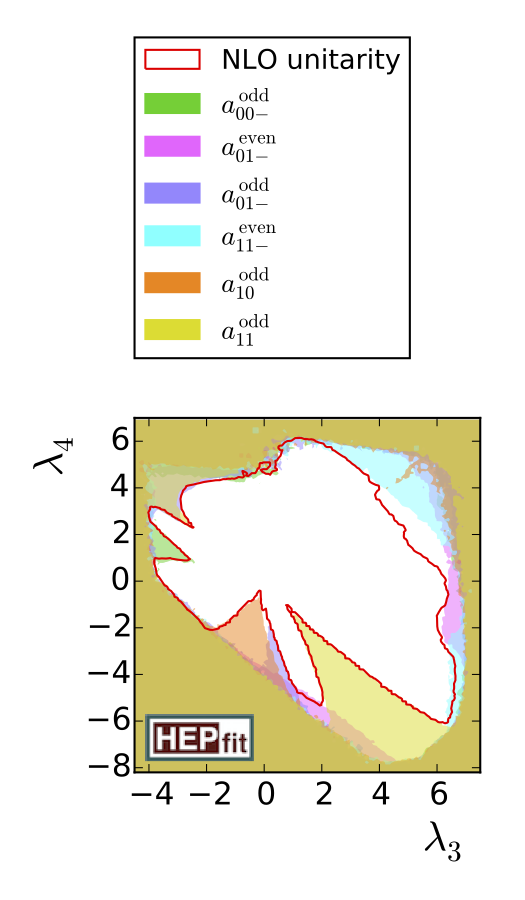}}
   \put(-33,217){\includegraphics[width=284pt,trim=0 250 0 0,clip=true]{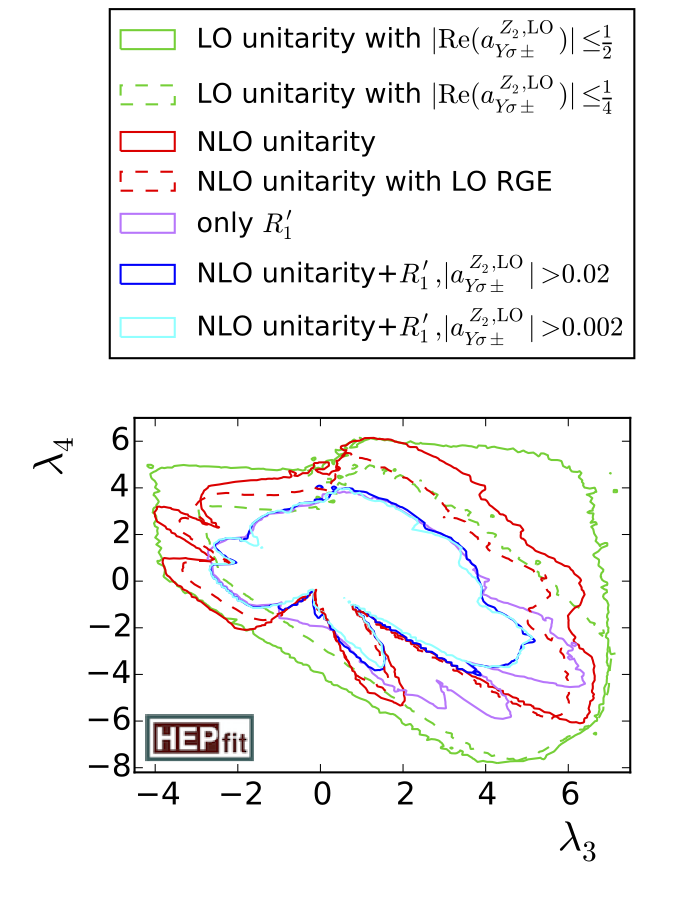}}
   \end{picture}
  \caption{Comparisons of the effects of different constraints in the $\lambda_4$ vs.~$\lambda_3$ plane. Left: The solid green, red, and blue curves have the same meaning here as the shaded regions of the same colour in Fig.~\ref{fig:lambdaswithunitarity}. The dashed green curve shows the effect of (arbitrarily) requiring the LO unitarity condition to be more restrictive. The pink curve demonstrates the impact of the perturbativity bounds without the unitarity bounds. The cyan curve requires $|a^{\mathbb{Z}_2,\text{LO}}_{Y\sigma\pm}| > 0.002$ in order for the perturbativity bounds to be enforced, rather than 0.02, leading to no significant change in the allowed parameter space. Right: Breakdown of the single effects of the unitarity constraints. Only the most constraining eigenvalues are displayed.}
  \label{fig:lambda3vslambda4}
\end{figure}

A closer look at different variations of our conditions explained in Sec.~\ref{sec:theoconstraints} is shown in the left panel of Fig.~\ref{fig:lambda3vslambda4} in the $\lambda_4$ vs.~$\lambda_3$ plane. The green, red and blue solid lines correspond to the contours of the same colour in the previous figure; all lines are the $99.7$\% probability boundaries.
As explained above, previous studies used $1/4$ rather than $1/2$ as upper limit for the real part of the LO unitarity eigenvalues. This choice is represented by the green dashed line and is almost always less stringent than $R_1^{\prime}$-perturbative NLO unitarity.
The red dashed line uses LO RGE instead of the NLO RGE which apply in all other cases. As already stated in Ref.~\cite{Chowdhury:2015yja}, the NLO RGE ``stabilize'' the potential with respect to the LO expressions in the sense that for the same starting point one runs into non-perturbative values for the quartic couplings at much lower scales with LO RGE. That is why larger values for the $\lambda_i$ are accessible at the electroweak scale if one uses the NLO RGE.
What happens if one only requires that $R_1^{\prime}<1$ without imposing NLO unitarity can be seen at the pink contour. In other words, the blue line should be the combination of the red (NLO unitarity) and the pink one. Only for $\lambda_3>4$, one can see that the combination of both sets of constraints is stronger than their individual impacts.
Finally, the cyan contour is the result of using a ten times smaller threshold for the LO part of $R_1^{\prime}$. However, compared to our typical limit of $0.02$ this is not substantially different.

Again in the $\lambda_4$ vs.~$\lambda_3$ plane, we also show the individual contributions of the relevant single NLO eigenvalues in the right panel of Fig.~\ref{fig:lambda3vslambda4}. The shaded areas are excluded at $99.7$\% probability by the eigenvalues indicated in the legend. It is worth noting that only ``$-$'' solutions seem to be important in this plane.

\begin{figure}
   \begin{picture}(450,600)(0,0)
   \put(0,0){\includegraphics[width=450pt]{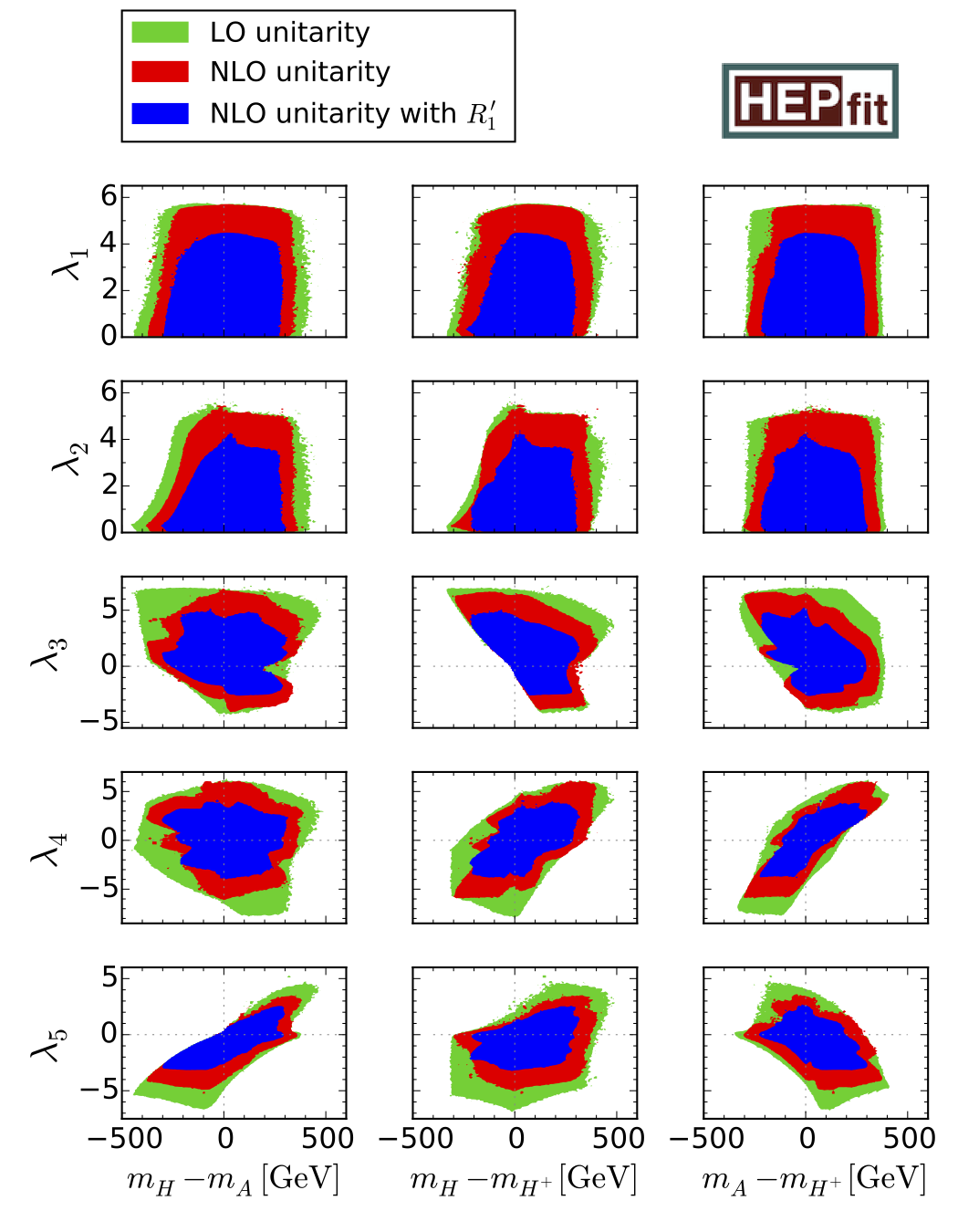}}
   \end{picture}
  \caption{$99.7\%$ probability regions in the $\lambda_i$ vs.~$(m_j - m_k)$ planes. The colours of the regions have the same meaning as those of Fig.~\ref{fig:lambdaswithunitarity}.}
  \label{fig:massdifferences}
\end{figure}

Going from the potential parametrisation to the physical parameters, one can see how the different constraints on the $\lambda_i$ couplings translate into restrictions on the mass differences between the heavy Higgs bosons $H$, $A$ and $H^+$ in Fig.~\ref{fig:massdifferences}. Like in Fig.~\ref{fig:lambdaswithunitarity}, we observe a hierarchy between LO unitarity, NLO unitarity and NLO unitarity with the perturbativity requirement $R_1^{\prime}<1$, with the first set of constraints being the weakest bound and the last being the strongest bound. While LO unitarity allows for maximal $|m_H-m_A|$, $|m_H-m_{H^+}|$ and $|m_A-m_{H^+}|$ of $500$ GeV, the perturbative NLO unitarity conditions sets upper limits on the mass splittings of around $360$ GeV. $m_{H^+}>m_H$ and $m_A>m_H$ are already almost excluded by LO unitarity for $\lambda_3<0$ and $\lambda_5>0$, respectively; we can see that after the inclusion of NLO unitarity with the $R_1^{\prime}$ condition also other constellations like $\lambda_3=0$ feature significantly smaller possible mass differences.

\begin{figure}
   \begin{picture}(450,450)(0,0)
    \put(0,0){\includegraphics[width=450pt]{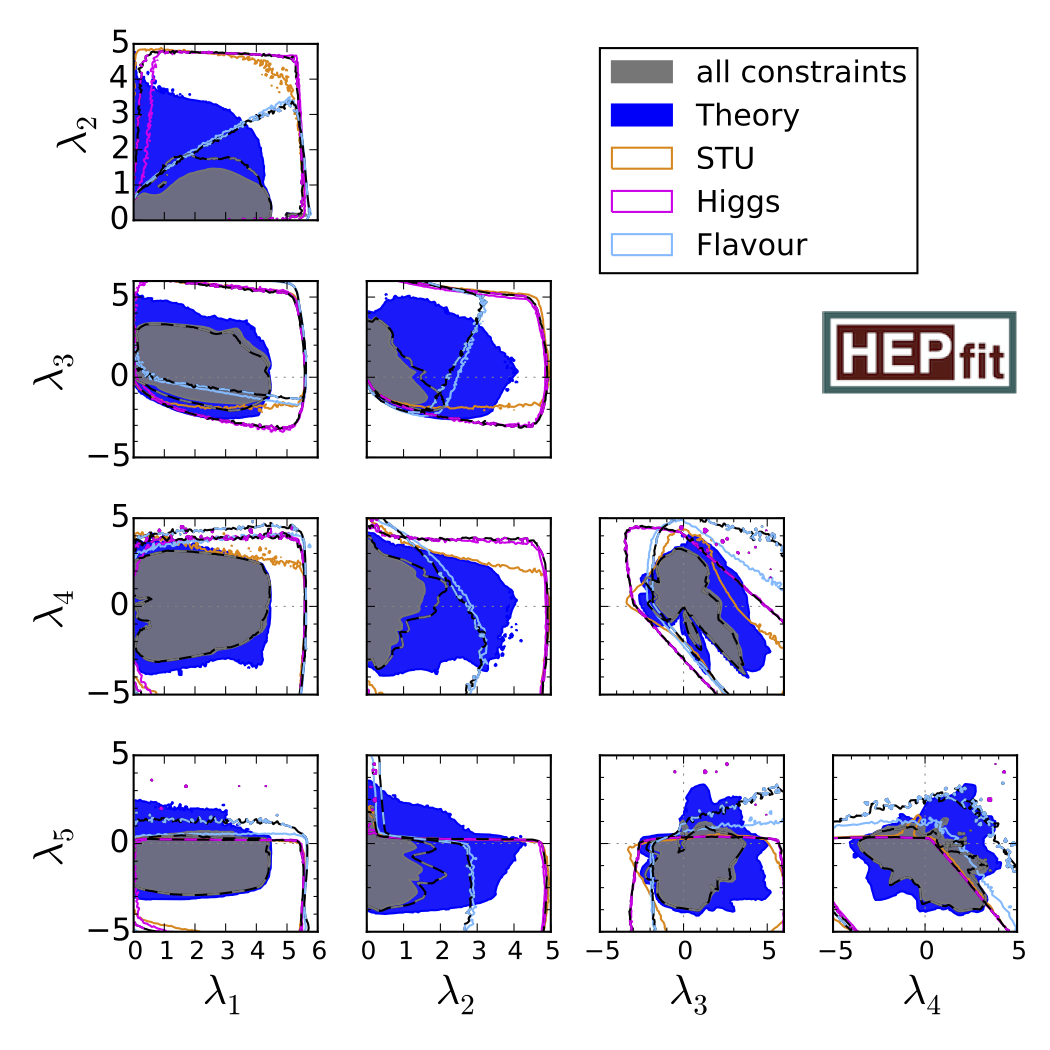}}
   \end{picture}
  \caption{$\lambda_i$ vs.~$\lambda_j$ planes including experimental constraints. The blue shaded regions are the same $99.7$\% probability areas as in Fig.~\ref{fig:lambdaswithunitarity}, while the orange, pink and light blue lines mark the $95.4$\% probability boundaries of fits using only the oblique parameters ($STU$), all direct Higgs observables (that is $h$ signal strengths and searches for $H$ and $A$) or the flavour observables $\Delta m_{B_s}$ and $\mathcal{B}(\bar{B}\to X_s\gamma)$. The grey contours are compatible with all theoretical and experimental bounds at a probability of $95.4$\%. The solid lines are understood as the type II contours, the coloured dashed lines represent the corresponding type I fits.}
  \label{fig:lambdaswithexp}
\end{figure}

Fig.~\ref{fig:lambdaswithexp} contains the same quartic coupling planes as Fig.~\ref{fig:lambdaswithunitarity}, but additionally the experimental data has been taken into account. The blue region survives all theoretical constraints as mentioned at the beginning of Section \ref{sec:theoconstraints} and is identical with the blue contours of the previous figures. The unfilled contours have been obtained using only one of the following three sets of inputs in combination with the requirement that the scalar potential is stable up to $750$ GeV: the oblique parameters (labelled ``STU''); $h$ signal strengths and $H$ and $A$ searches (``Higgs''); $\Delta m_{B_s}$ and $\mathcal{B}(\bar{B}\to X_s\gamma)$ (``Flavour''). In both types, the first set is most constraining for negative $\lambda_3$ or positive $\lambda_4$, the second set excludes $\lambda_5>0.4$, and the third set of inputs yields $\lambda_2<3.5$ and $\lambda_3>-2$.
Finally, the grey regions denote the combination of all theoretical and experimental constraints in type II and the grey dashed lines correspond to the type I fits. The allowed $\lambda_1$ and $\lambda_4$ intervals are similar to the ones obtained in Fig.~\ref{fig:lambdaswithunitarity}, but the other three receive significant additional restrictions from the experimental bounds: with a probability of $95.4$\%, $\lambda_2$ cannot exceed $1.6$ ($1.2$), $\lambda_3$ has to be within $-1.6$ and $3.0$ ($-1.3$ and $3.1$) and the allowed $\lambda_5$ interval is between $-2.7$ and $0.3$ ($-2.7$ and $0.5$), when marginalizing over all parameters in type I (type II).

\begin{figure}
   \begin{picture}(450,300)(0,0)
    \put(0,0){\includegraphics[width=450pt]{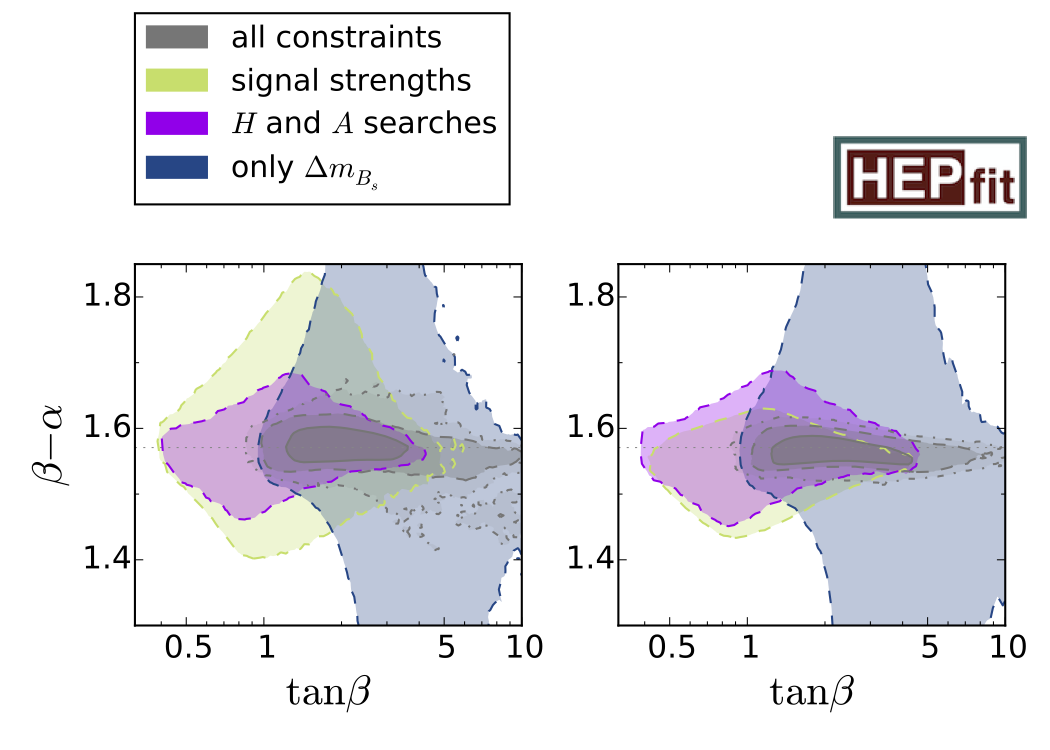}}
   \end{picture}
  \caption{The $\beta-\alpha$ vs. $\tan \beta$ planes of type I (left) and type II (right) with the single contributions of the most important constraints: the $h$ signal strengths in light green, the heavy neutral Higgs searches in pink and the mass difference between the $B_s$ and $\bar{B}_s$ mesons in dark blue; the grey contours stem from the combined fit to all constraints. The dashed lines represent the $95.4$\% probability boundaries, the grey solid and dash-dotted ones the $68.3$\% and $99.7$\% contours, respectively. The grey dotted line indicates the alignment limit $\beta-\alpha=\pi/2$.}
  \label{fig:tanbbma}
\end{figure}

Again turning towards the physical parameters, we show the allowed parameter space in the $\beta-\alpha$ vs. $\tan \beta$ plane for type I and II in the left and right panels of Fig.~\ref{fig:tanbbma}, respectively. The most important bounds in this plane comes from the $h$ signal strengths, the heavy Higgs searches and $\Delta m_{B_s}$; their $95.4$\% bounds are also depicted individually. The Higgs observables strongly constrain the difference between $\alpha$ and $\beta$. In the final fit with all constraints, the deviation from the alignment limit, $|\beta-\alpha-\pi/2|$, cannot exceed 0.15 and 0.04 in type I and II, respectively. (This corresponds to a maximal deviation of $\sin(\beta-\alpha)$ from $1$ by $0.01$ and $7\cdot 10^{-4}$, respectively.) The mass difference in the $B_s$ system sets a type independent lower bound on $\beta$ for the chosen mass priors. More details about the effects of the signal strengths can be found in Fig.s~\ref{fig:tanbvsbma_typeI} and \ref{fig:tanbvsbma} in \hyperref[sec:appendixA]{Appendix A}.

\begin{figure}
   \begin{picture}(450,600)(0,0)
    \put(0,0){\includegraphics[width=450pt]{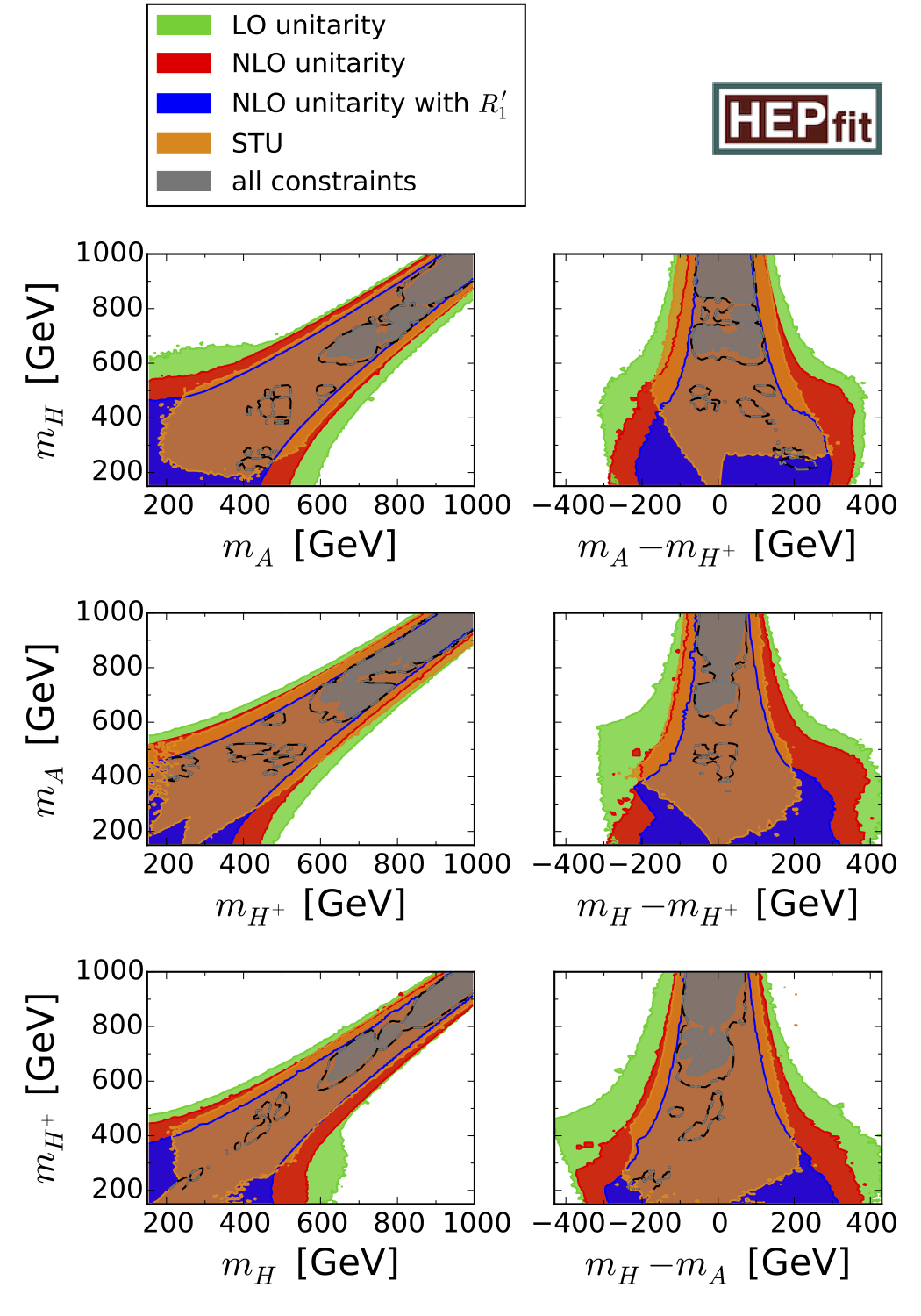}}
   \end{picture}
  \caption{Allowed regions in the heavy Higgs boson masses and their mass differences planes in the 2HDM of type I (dashed lines) and type II (solid lines). The unitarity bounds to the green, red and blue regions are meant at a probability of $99.7$\%, and the orange and grey lines mark the $95.4$\% boundaries.}
  \label{fig:masses_typeII}
\end{figure}

In Fig.~\ref{fig:masses_typeII}, we plot the allowed ranges for the heavy Higgs boson masses and their mass differences after imposing the theoretical and experimental constraints for type I and type II. The green, red and blue regions depict the $99.7$\% allowed parameter space for the various unitarity conditions discussed above. The orange region is the allowed by the $STU$ observables at $95.4$\%. Finally, the grey region is the available parameter space after all the theoretical and experimental constraints are taken into consideration.
Even if the perturbative NLO unitarity contour represents a larger probability boundary than the oblique parameter contour ($99.7$\% for the former, $95.4$\%), it is more stringent for masses above $400$ GeV and thus the dominant constraint in the high mass regime. It allows for maximal mass splittings between $m_H$, $m_A$ and $m_{H^+}$ of around $100$ GeV for masses above $600$ GeV. After the inclusion of the LHC searches for heavy neutral Higgs bosons, we observe that the remaining parameter space is disconnected. The largest gap occurs around $m_{H,A}\approx 550$ GeV. The reason for this discontinuity is that our fits are incompatible with the observed ATLAS and CMS diphoton cross sections around this mass. For details, we refer to Fig.s~\ref{fig:massesandangles_typeI} and \ref{fig:massesandangles} in Appendix \ref{sec:appendixA}.
With a probability of $95.4$\%, $H$ and $H^+$ can be as light as $210$ GeV, and $m_A$ cannot be smaller than approximately $400$ GeV in type I.
In type II, masses below $600$ GeV are excluded at $95.4$\% after the inclusion of $\mathcal{B}(\bar{B}\to X_s\gamma)$ to the fit.

\begin{figure}
   \begin{picture}(450,550)(0,0)
   \put(0,0){\includegraphics[width=450pt]{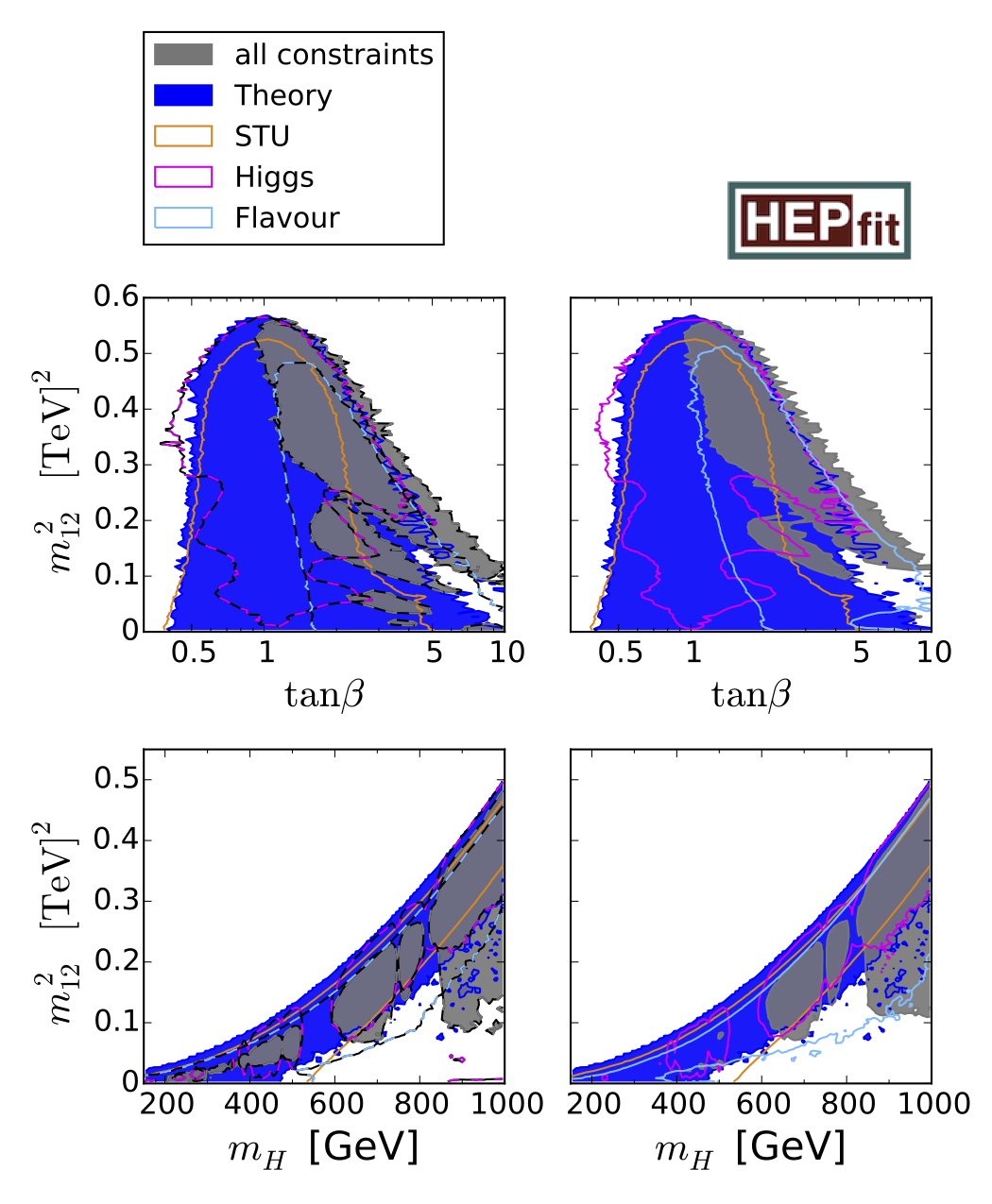}}
   \end{picture}
  \caption{$m_{12}^2$ vs. $\tan \beta$ and $m_{12}^2$ vs. $m_H$ planes in type I (left) and type II (right). The colours have the same meaning than in Fig.~\ref{fig:lambdaswithexp} with the difference that also the grey type I contour was filled here.}
  \label{fig:m12sq}
\end{figure}

Finally, we address the soft $\mathbb{Z}_2$ symmetry breaking parameter $m_{12}^2$. In Fig.~\ref{fig:m12sq} we show its dependence on $\tan \beta$ and the $H$ Higgs mass in the two discussed types. While for the theoretical set of constraints a strong correlation between the heavy Higgs mass and $m_{12}^2$ is visible, this gets somewhat relaxed if one adds experimental data to the fit. This is due to the flavour constraints, which favour larger $\tan \beta$ and $m_{H^+}$ values. The most important result here is that an unbroken $\mathbb{Z}_2$ symmetry can be excluded with a probability of $95.4$\% in the combined fit to the type II; the single sets of constraints are individually compatible with an exact $\mathbb{Z}_2$ symmetry. The lowest $95.4$\% allowed value for $m_{12}^2$ is ($370$ GeV)$^2$, if we marginalize over all other parameters.

\section{Conclusions}
\label{sec:conclusions}

The determination of the NLO unitarity constraints to the 2HDM with a softly broken $\mathbb{Z}_2$ symmetry mitigates the problem of how to tame higher order contributions involving large quartic couplings. The expressions have been derived in Ref.~\cite{Grinstein:2015rtl}, and in this article we perform the first general fits to them in the 2HDM of type I and II, making use of the publicly available package \HEPfit. One important result is that wavefunction renormalization contributions can be safely neglected in these models.
In our fits we also apply the suppression of non-perturbative higher order contributions with the $R_1^{\prime}$ condition, requiring that the NLO part cannot be larger in magnitude than the LO contribution if the latter is not accidentally small. We find that both steps, going from LO to NLO unitarity and comparing NLO unitarity with $R_1^{\prime}$-perturbative NLO unitarity, individually put strong bounds on the 2HDM parameters. If we add all other relevant theoretical constraints, that is stability and positivity of the scalar potential up to a scale of $750$ GeV, the quartic $\lambda_i$ couplings cannot exceed $5.8$ in magnitude and the mass differences between $m_H$, $m_A$ and $m_{H^+}$ cannot be larger than approximately $360$ GeV. (The latter even reduces to maximally $100$ GeV for heavy Higgs masses above $800$ GeV.) To our knowledge, this currently represents the strongest reliable bound on the mass splittings.

As a next step, we have added all the relevant experimental constraints to the fit: the electroweak precision data in form of the oblique parameters, the complete set of LHC Run I results and the most important flavour observables. These bounds constrain the quartic couplings even further: the allowed intervals for the quartic couplings are 

\begin{tabular*}{250pt}{rlrlrlrlrl}
\hspace*{-14pt} $0\leq$ & $\lambda_1<4.2$, & \quad $0\leq$ & $\lambda_2<1.6$, & \quad $-1.6<$ & $\lambda_3<3.0$, & \quad$-2.5<$ & $\lambda_4<2.9$, & \quad $-2.7<$ & $\lambda_5<0.3$\\[1pt]
\end{tabular*}

\noindent in type I and

\begin{tabular*}{250pt}{rlrlrlrlrl}
\hspace*{-14pt} $0\leq$ & $\lambda_1<4.2$, & \quad $0\leq$ & $\lambda_2<1.2$, & \quad $-1.3<$ & $\lambda_3<3.1$, & \quad$-2.5<$ & $\lambda_4<2.9$, & \quad $-2.7<$ & $\lambda_5<0.5$\\[1pt]
\end{tabular*}

\noindent in type II with a probability of $95.4$\%. For the physical parameters, we find that $\tan \beta$ cannot be smaller than $1$ in both discussed types of the 2HDM. The deviation from the alignment limit $|\beta-\alpha-\pi/2|$ cannot exceed 0.15 (0.04) in type I (type II).
In type I the global fit produces lower $95.4$\% bounds of $210$ GeV for $m_H$ and $m_{H^+}$ and $410$ GeV for $m_A$, while these limits are around $650$ GeV for all three heavy Higgs masses in type II. In the latter case, also an unbroken $\mathbb{Z}_2$ symmetry can be ruled out at $95.4$\%; the soft $\mathbb{Z}_2$ breaking parameter $m_{12}^2$ has to be larger than ($370$ GeV)$^2$.

\begin{acknowledgments}
We thank Marco Fedele, Enrico Franco, Benjam\'{i}n Grinstein, Ayan Paul, Maurizio Pierini, Luca Silvestrini, and Patipan Uttayarat for useful discussions. The research leading to these results has received funding from the European Research Council under the European Union's Seventh Framework Programme (FP/2007-2013) / ERC Grant Agreement n.~279972. This work was supported in part by the MIUR-FIRB under grant no.~RBFR12H1MW.
\end{acknowledgments}

\section*{Appendix A}
\label{sec:appendixA}

\begin{figure}
   \begin{picture}(450,600)(0,0)
    \put(0,0){\includegraphics[width=450pt]{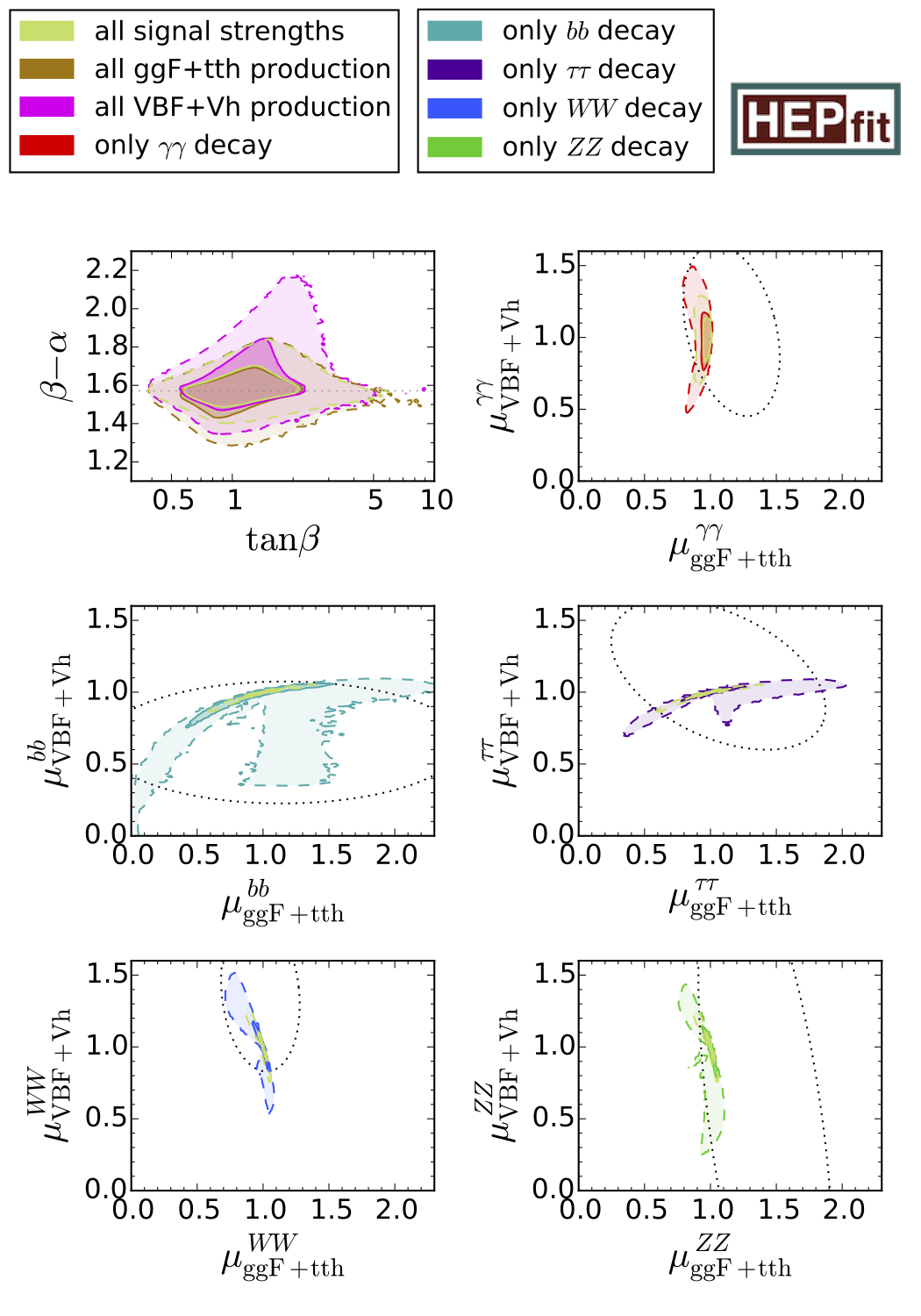}}
   \end{picture}
  \caption{68.3{\%} (solid) and 95.4{\%} (dashed) regions in the  $\beta-\alpha$ vs. $\tan \beta$ plane and different signal strengths in type I. The colours of the single decay channels match the ones chosen for the official combination of ATLAS and CMS signal strengths \cite{Khachatryan:2016vau}, which are also approximated by the black dotted ellipses. In the top left panel, we also mark the alignment limit $\beta-\alpha=\pi/2$ by a grey dotted line.}
  \label{fig:tanbvsbma_typeI}
\end{figure}

\begin{figure}
   \begin{picture}(450,600)(0,0)
    \put(0,0){\includegraphics[width=450pt]{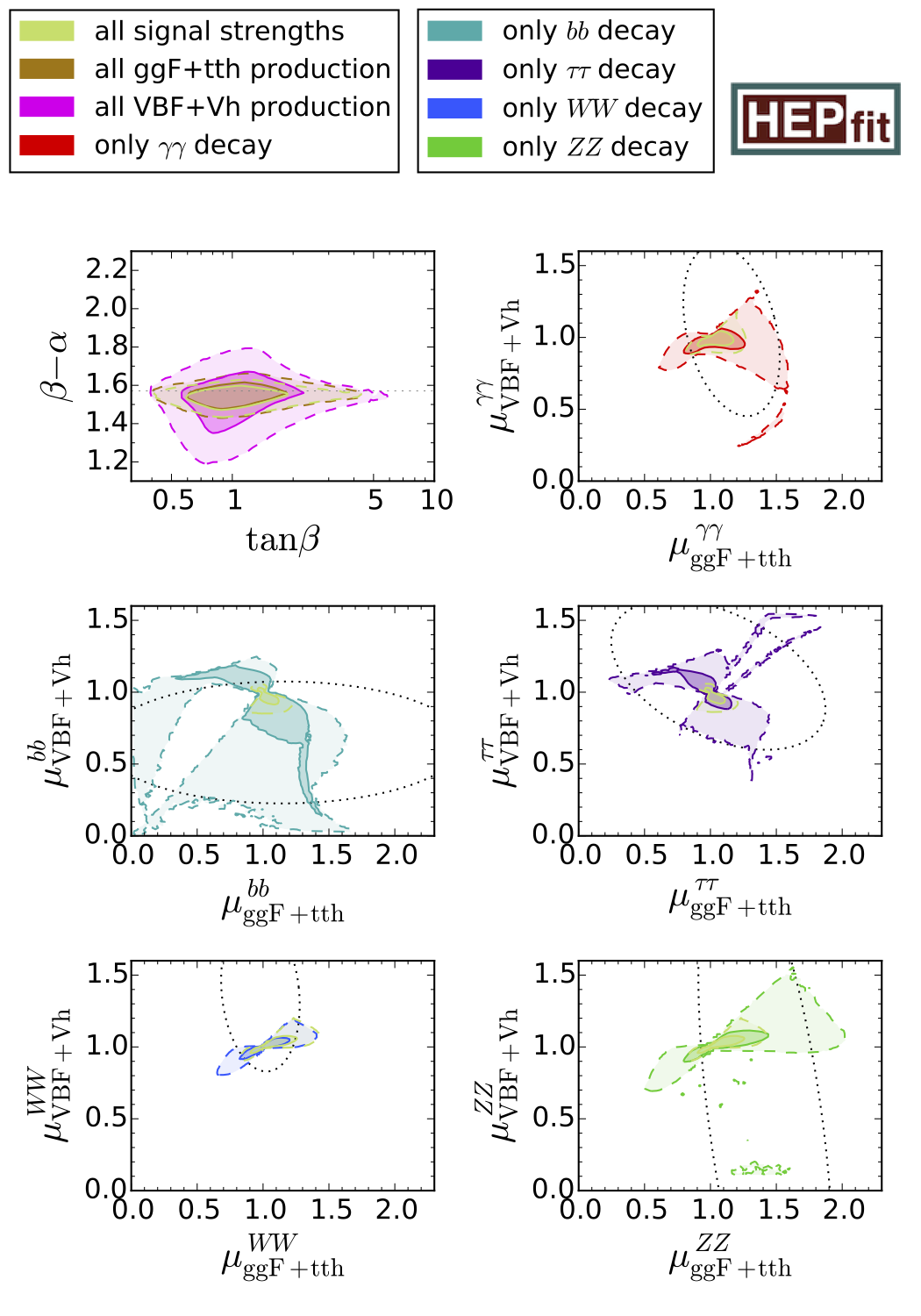}}
   \end{picture}
  \caption{Same as Fig.~\ref{fig:tanbvsbma_typeI} but for type II.}
  \label{fig:tanbvsbma}
\end{figure}

\begin{figure}
   \begin{picture}(450,600)(0,0)
    \put(0,0){\includegraphics[width=450pt]{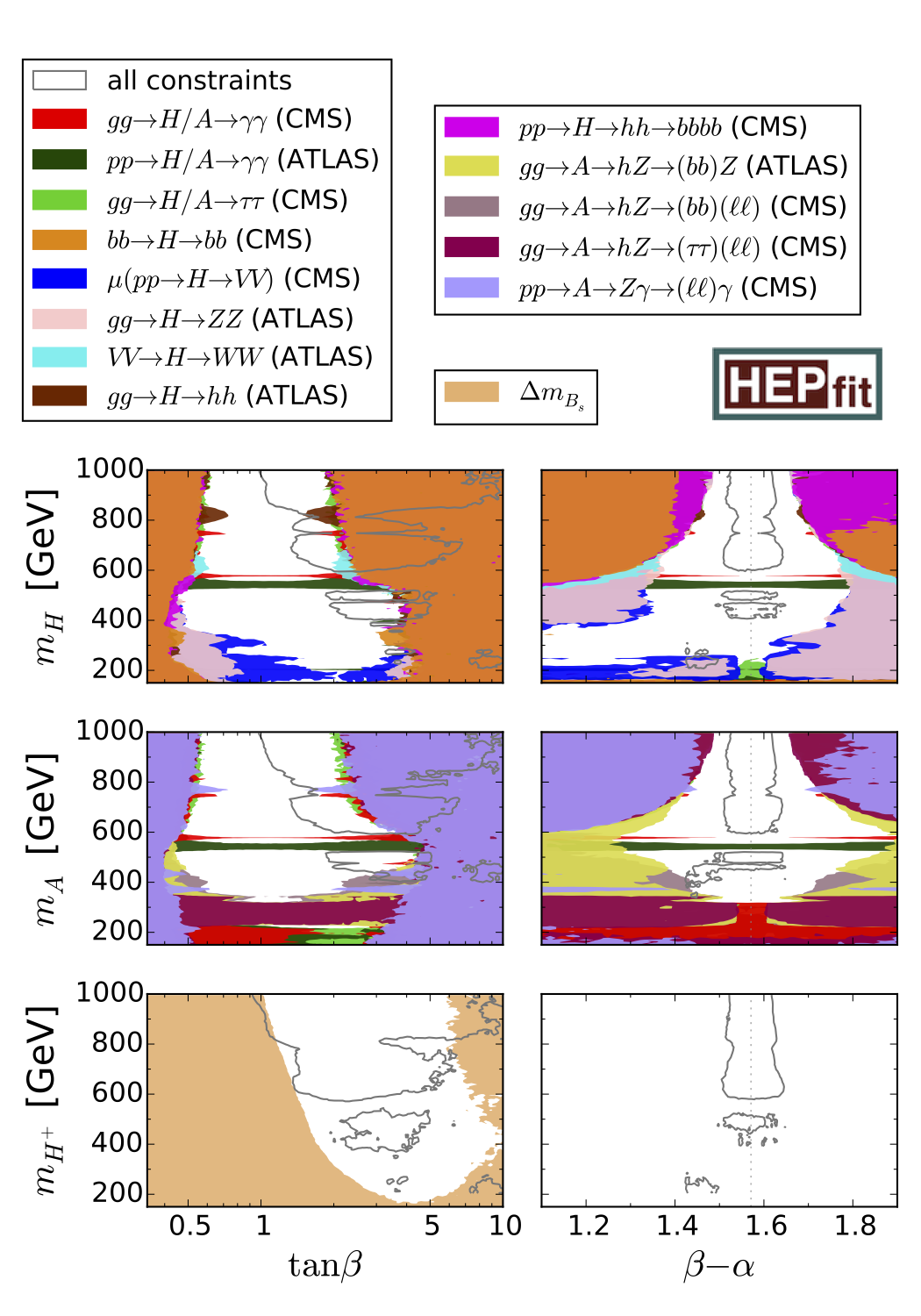}}
   \end{picture}
  \caption{Exclusion boundaries for type I fits in the heavy Higgs mass vs. $\tan\beta$ (left column) and heavy Higgs mass vs. $\beta-\alpha$ (right column) at $95.4\%$ probability. The grey contour denotes the available parameter space at $95.4\%$ probability, after imposing all the theoretical and experimental constraints.}
  \label{fig:massesandangles_typeI}
\end{figure}

\begin{figure}
   \begin{picture}(450,600)(0,0)
    \put(0,0){\includegraphics[width=450pt]{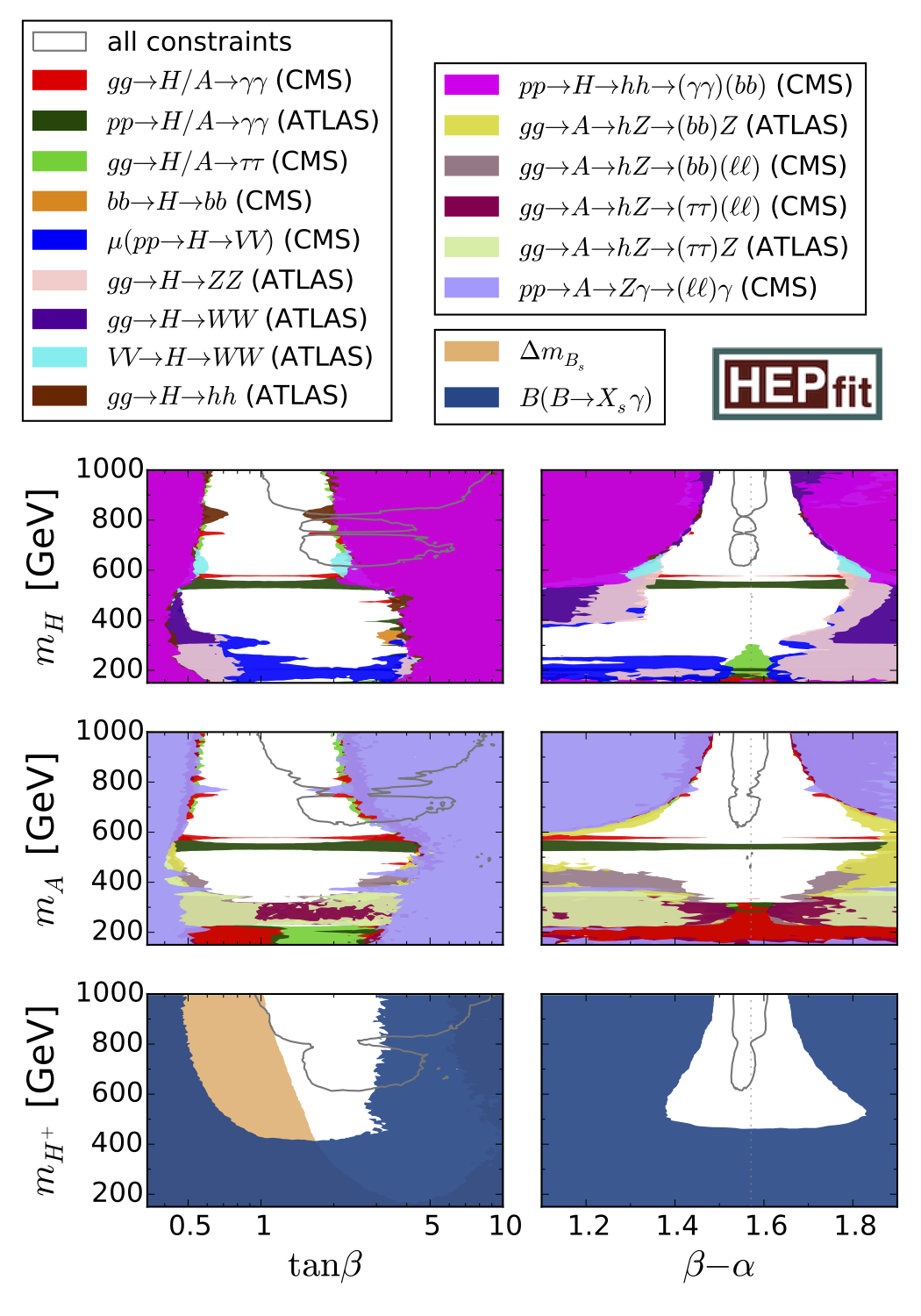}}
   \end{picture}
  \caption{Like Fig.~\ref{fig:massesandangles_typeI} but for type II fits.}
  \label{fig:massesandangles}
\end{figure}

\begin{figure}
   \begin{picture}(450,450)(0,0)
    \put(0,0){\includegraphics[width=450pt]{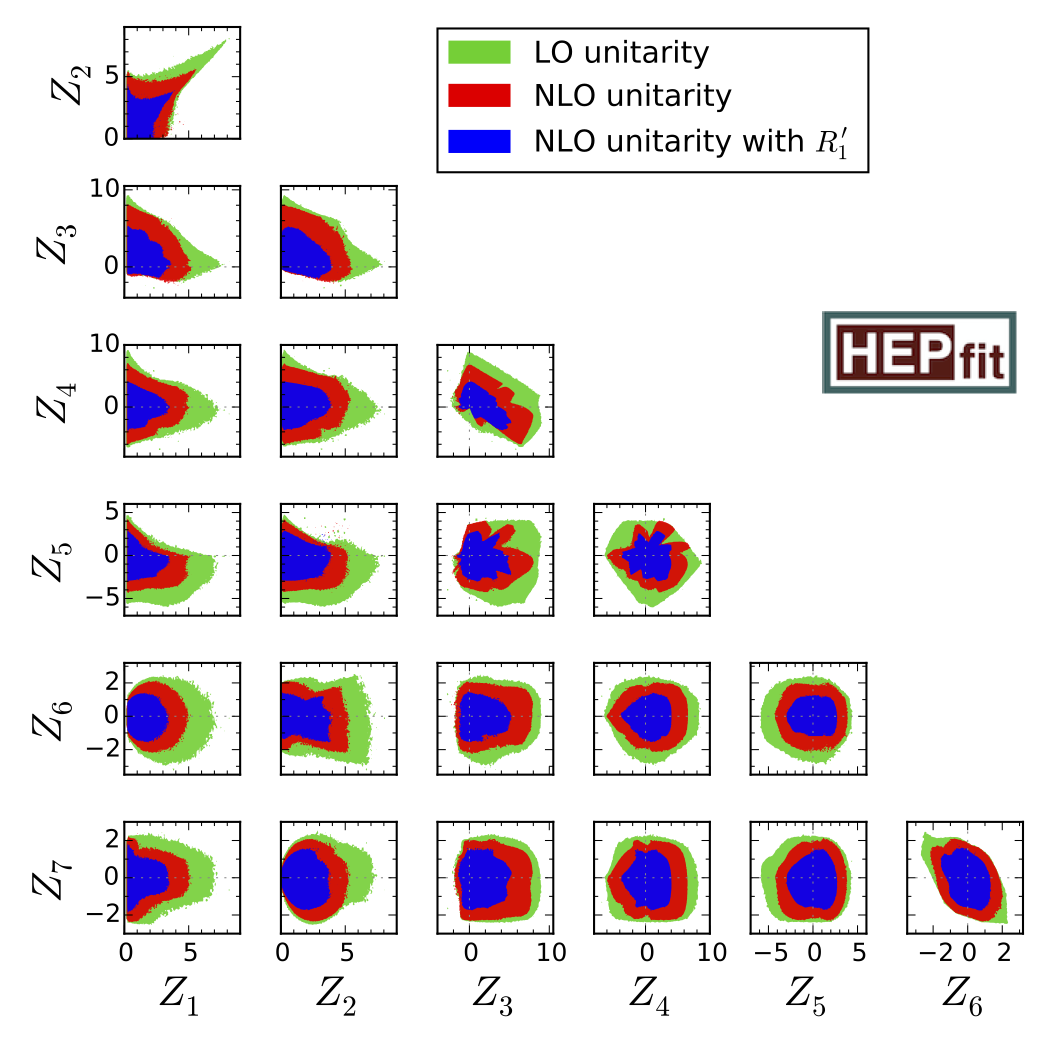}}
   \end{picture}
  \caption{Planes of the quartic couplings $Z_i$ of the Higgs basis parametrisation from \eqref{eq:potHiggsBasis}. The colours are analogous to Fig.~\ref{fig:lambdaswithunitarity}.}
  \label{fig:Zplanes}
\end{figure}

\begin{figure}
   \begin{picture}(450,450)(0,0)
    \put(0,0){\includegraphics[width=450pt]{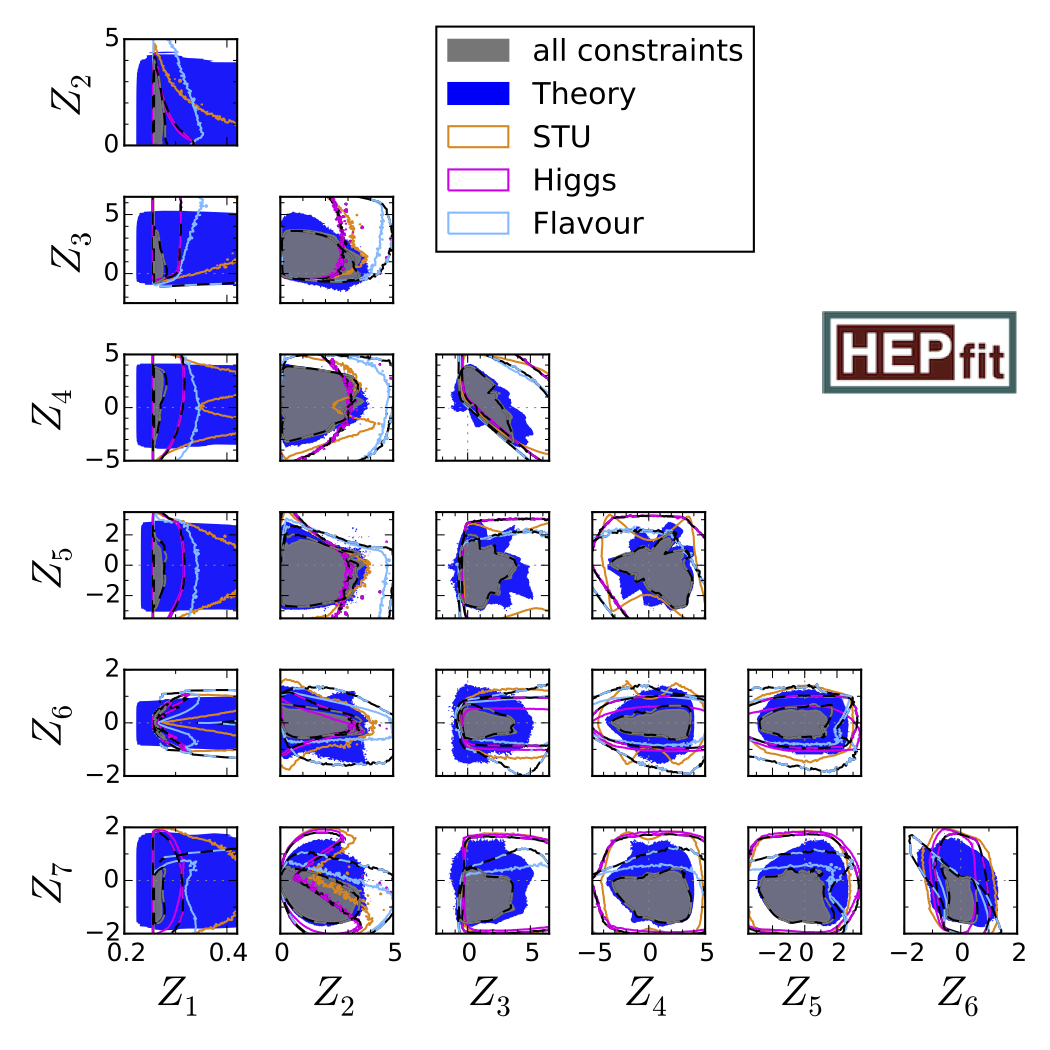}}
   \end{picture}
  \caption{Planes of the quartic couplings $Z_i$ of the Higgs basis parametrisation from \eqref{eq:potHiggsBasis} including experimental inputs. The colours are analogous to Fig.~\ref{fig:lambdaswithexp}.}
  \label{fig:Zplanes_exp}
\end{figure}

In this appendix, we present supplementary figures of the 2HDM parameter space: dedicated fits of the different signal strength measurements and $H$ and $A$ searches in type I and II in Fig.s~\ref{fig:tanbvsbma_typeI} to \ref{fig:massesandangles}, and the quartic couplings of the so-called Higgs basis in Fig.s~\ref{fig:Zplanes} and \ref{fig:Zplanes_exp}.

In Fig.~\ref{fig:tanbvsbma_typeI}, we show the effect of the $h$ signal strengths on the $\tan\beta$ vs.~$\beta-\alpha$ plane for 2HDM of type I. In the top left panel of Fig.~\ref{fig:tanbvsbma_typeI}, the effect of considering all the five signal strengths in the ``ggF+tth'' production modes on this plane is represented by the orange shaded region, considering all the five signal strengths in the ``VBF+VH'' production modes are shown in the pink region, and the light green shaded region depicts the allowed parameter space when all the ten signal strengths are taken into consideration. In order to compare the latter with the effect of each of the five decay modes individually, we separately plot the single decay modes at a time on the rest of the panels of Fig.~\ref{fig:tanbvsbma_typeI} overlaid with the fit with all signal strengths. For each of these additional panels we also indicate the latest 8 TeV signal strength correlation contours at 68\% CL taken from Ref.~\cite{Khachatryan:2016vau}. In all the panels, the filled regions with solid (dashed) lines represent the 68.3\% (95.4\%) probability contours as obtained from the fits. Fig.~\ref{fig:tanbvsbma} displays the same panels as Fig.~\ref{fig:tanbvsbma_typeI} but for type II.

Fig.s~\ref{fig:massesandangles_typeI} and \ref{fig:massesandangles} compare the most important constraints on the heavy Higgs masses vs. $\tan\beta$ planes (left column) and on the heavy Higgs masses vs. $\beta-\alpha$ planes (right column) in type I and type II, respectively. For the first four panels, the relevant $H$ (top row) and $A$ (second row) searches are represented by the shaded regions, which they exclude. For the attribution of the colours, we refer to the legends.
The left panel of the bottom row of Fig.s~\ref{fig:massesandangles_typeI} and \ref{fig:massesandangles} shows in dark blue the constraint from the mass difference in the $B_s$ system on the charged Higgs mass vs. $\tan\beta$ plane, which excludes $\tan \beta<1$ for the chosen $m_{H^+}$ range. Fig.~\ref{fig:massesandangles} additionally features the constraint from $\mathcal{B}(\bar{B}\to X_s\gamma)$ disfavouring charged Higgs masses below $410$ GeV. The grey contours in all the panels of Fig.s~\ref{fig:massesandangles_typeI} and \ref{fig:massesandangles} depict the allowed parameter space after all of the theoretical and experimental constraint have been taken into account. All contours represent the $95.4$\% probability boundaries.
In the $m_H$ and $m_A$ planes, the searches for neutral Higgs particles mainly disfavour very small and very large values of $\tan \beta$. Around $550$ GeV even all $\tan \beta$ values are incompatible with the measured diphoton events at ATLAS and CMS. In the data of both collaborations, the observed upper limits on $\sigma \mathcal{B}$ are significantly larger than the expected exclusion limits at this invariant mass; an excess which cannot be explained in the context of a 2HDM of type I or II. The grey contours reflect the features of all important constraints as well as their interplay: also in the $m_{H/A}$ vs. $\tan \beta$ planes, small $\tan \beta$ values are excluded in the fit with all constraints, because the masses of the neutral Higgs bosons cannot be very different from the $H^+$ mass due to unitarity (see Fig.~\ref{fig:masses_typeII}).

Instead of the general parametrisation of the potential in \eqref{eq:pot}, one is free to choose a basis in which only one of the two transformed doublets, $H_1$ and $H_2$, obtain a VEV. This basis is called the Higgs basis \cite{Branco:1999fs,Davidson:2005cw}, and its potential can be written as
\begin{align}
 V
 &=Y_1 H_1^\dagger H_1^{\phantom{\dagger}}
   +Y_2 H_2^\dagger H_2^{\phantom{\dagger}}
   +Y_3 (  H_1^\dagger H_2^{\phantom{\dagger}}
              + H_2^\dagger H_1^{\phantom{\dagger}})
   +\tfrac12 Z_1( H_1^\dagger H_1^{\phantom{\dagger}})^2
   +\tfrac12 Z_2( H_2^\dagger H_2^{\phantom{\dagger}})^2
 \nonumber \\
 &\phantom{{}={}}
  +Z_3( H_1^\dagger H_1^{\phantom{\dagger}})
            ( H_2^\dagger H_2^{\phantom{\dagger}})
  +Z_4( H_1^\dagger H_2^{\phantom{\dagger}})
            ( H_2^\dagger H_1^{\phantom{\dagger}})
  +\tfrac12 Z_5 \left[ ( H_1^\dagger H_2^{\phantom{\dagger}})^2
                      +( H_2^\dagger H_1^{\phantom{\dagger}})^2 \right]
 \nonumber \\
 &\phantom{{}={}}
 +\left[ Z_6 ( H_1^\dagger H_1^{\phantom{\dagger}})
           +Z_7 ( H_2^\dagger H_2^{\phantom{\dagger}}) \right] 
    (  H_1^\dagger H_2^{\phantom{\dagger}}
              + H_2^\dagger H_1^{\phantom{\dagger}}). \label{eq:potHiggsBasis}
\end{align}

Only five of the seven quartic couplings $Z_i$ are linearly independent. One can see from Fig.~\ref{fig:Zplanes} that they get constrained by the different unitarity conditions in a similar way than the $\lambda_i$ in Fig.~\ref{fig:lambdaswithunitarity}, with the $R_1^{\prime}$-perturbative NLO expressions being stronger than simple NLO unitarity, which itself is an improvement of LO unitarity. While the latter does not allow for $|Z_i|>9$, NLO unitarity (with $R_1^{\prime}$) sets upper limits of approximately $8$ ($5$) on the absolute values of the $Z_i$. Analogous to Fig.~\ref{fig:lambdaswithexp}, we also show the impact of the experimental constraints on the $Z_i$ vs. $Z_j$ planes in Fig.~\ref{fig:Zplanes_exp}. Especially $Z_1$, $Z_6$ and $Z_7$ suffer strong additional restrictions from the experiments.

\section*{Appendix B}
\label{sec:appendixB}
For the reader's convenience we list the minimal set of elements of the matrix $\mathbf{a}_0$ needed to write its eigenvalues, $a_0$, at next-to-leading order accuracy in the limit that the wavefunction renormalization contribution is neglected. In what follows, each $B_N$ corresponds in this approximation to Eq.~(B.N) of Appendix~B of Ref.~\cite{Grinstein:2015rtl}. The complete expressions for $\mathbf{a}_0$ can be found in Appendices~B and~C of Ref.~\cite{Grinstein:2015rtl}.
\begin{align*}
B_1 &= - 3 \lambda_1 + \frac92 \beta_{\lambda_1} + \frac{1}{16 \pi^2}\left(i \pi - 1\right) \left(9 \lambda_1^2 + \left(2 \lambda_3 + \lambda_4\right)^2\right) \\
B_2 &= - 3 \lambda_2 + \frac{9}{2} \beta_{\lambda_2} + \frac{1}{16\pi^2} \left(i \pi - 1\right) \left(9 \lambda_2^2 + \left(2 \lambda_3 + \lambda_4\right)^2\right) \\
B_3 &= - \left(2 \lambda_3 + \lambda_4\right) + \frac32 \left(2 \beta_{\lambda_3} +\beta_{\lambda_4}\right) + \frac{3}{16\pi^2} \left(i \pi - 1\right) \left(\lambda_1 + \lambda_2\right) \left(2 \lambda_3 + \lambda_4\right) \\
B_4 &= - \left(\lambda_3 + 2 \lambda_4\right) + \frac32 \left(\beta_{\lambda_3} + 2 \beta_{\lambda_4}\right) + \frac{1}{16\pi^2} \left(i \pi - 1\right) \left(\lambda_3^2 + 4 \lambda_3 \lambda_4 + 4\lambda_4^2 + 9 \lambda_5^2\right)\\
B_6 &= - 3 \lambda_5 + \frac92 \beta_{\lambda_5} + \frac{6}{16\pi^2} \left(i \pi -1\right) \left(\lambda_3 + 2 \lambda_4\right) \lambda_5  \\
B_7 &= - \lambda_1 + \frac32 \beta_{\lambda_1} + \frac{1}{16\pi^2} \left(i \pi -1\right) \left(\lambda_1^2 + \lambda_4^2\right)  \\
B_8 &= - \lambda_2 + \frac32 \beta_{\lambda_2} + \frac{1}{16\pi^2} \left(i \pi -1\right) \left(\lambda_2^2 + \lambda_4^2\right)  \\
B_9 &= - \lambda_4 + \frac32 \beta_{\lambda_4} + \frac{1}{16\pi^2} \left(i \pi -1\right) \left(\lambda_1 + \lambda_2\right) \lambda_4 \\
B_{13} &= - \lambda_3 +\frac32 \beta_{\lambda_3} + \frac{1}{16\pi^2} \left(i \pi -1\right)\left(\lambda_3^2 + \lambda_5^2\right)  \\
B_{15} &= - \lambda_5 +\frac32 \beta_{\lambda_5} + \frac{2}{16\pi^2} \left(i \pi - 1\right) \lambda_3 \lambda_5 \\
B_{19} &= - \left(\lambda_3 - \lambda_4\right) + \frac32 \left(\beta_{\lambda_3} - \beta_{\lambda_4}\right) + \frac{1}{16\pi^2} \left(i \pi -1\right)\left(\lambda_3 - \lambda_4\right)^2  \\
B_{20} &= - \lambda_1 + \frac32 \beta_{\lambda_1} + \frac{1}{16\pi^2} \left(i \pi -1\right) \left(\lambda_1^2 + \lambda_5^2\right)  \\
B_{21} &= - \lambda_2 + \frac32 \beta_{\lambda_2} + \frac{1}{16\pi^2} \left(i \pi -1\right) \left(\lambda_2^2 + \lambda_5^2\right) \\
B_{22} &= - \lambda_5 + \frac32 \beta_{\lambda_5} + \frac{1}{16\pi^2} \left(i \pi -1\right)\left(\lambda_1 + \lambda_2\right) \lambda_5 \\
B_{30} &= - \left(\lambda_3 + \lambda_4\right) + \frac32 \left(\beta_{\lambda_3} + \beta_{\lambda_4}\right) + \frac{1}{16\pi^2} \left(i \pi -1\right)\left(\lambda_3 + \lambda_4\right)^2 
\end{align*}
For completeness, the leading terms of the beta functions appearing in the above equations are,
\begin{align*}
16 \pi^2 \beta_{\lambda_1} &= 12 \lambda_1^2 + 4 \lambda_3^2 +4 \lambda_3 \lambda_4 +2 \lambda_4^2 +2 \lambda_5^2, \\
16 \pi^2 \beta_{\lambda_2} &= 12 \lambda_2^2 + 4 \lambda_3^2 +4 \lambda_3 \lambda_4 +2 \lambda_4^2 +2 \lambda_5^2, \\
16 \pi^2 \beta_{\lambda_3} &= 4 \lambda_3^2 +2 \lambda_4^2 + \left(\lambda_1 + \lambda_2\right)\left(6 \lambda_3 +2 \lambda_4\right) +2 \lambda_5^2, \\
16 \pi^2 \beta_{\lambda_4} &= \left(2 \lambda_1 +2 \lambda_2 +8 \lambda_3\right) \lambda_4 +4 \lambda_4^2 +8 \lambda_5^2, \\
16 \pi^2 \beta_{\lambda_5} &= \left(2 \lambda_1 +2 \lambda_2 +8 \lambda_3 +12 \lambda_4\right) \lambda_5.
\end{align*}
It is worth mentioning that here only the LO expressions for the $\beta$ functions should be used in order to be consistent with the order of perturbation theory. For the running in the fits we apply NLO RGE.

\section*{Appendix C}
\label{sec:appendixC}

In the Tables \ref{tab:STU} to \ref{tab:Flavour} we list all used experimental inputs for our fits with their corresponding references.

\begin{table}
\begin{tabular}{| c | c | c c c |}
\hline
\textbf{Pseudo-observable} & \textbf{Value} & \multicolumn{3}{|l|}{\textbf{Correlation matrix}}\\
\hline
$S$ & $0.09\pm 0.10$  & 1 & $0.86$ & $-0.54$ \\
$T$ & $0.10\pm 0.12$  & $0.86$ & 1 & $-0.81$ \\
$U$ & $0.01\pm 0.09$  & $-0.54$ & $-0.81$ & 1 \\
\hline
\end{tabular}
\caption{$S$, $T$, and $U$ values and correlations from \cite{deBlas:2016ojx}.}
\label{tab:STU}
\end{table}

\begin{table}
\begin{tabular}{| c | c | c c |}
\hline
\textbf{Signal strength} & \textbf{Value} & \multicolumn{2}{|l|}{\textbf{Correlation matrix}}\\
\hline
$\mu_\text{ggF+tth}^{\gamma \gamma}$ & $1.16\pm 0.26$  & 1 & $-0.30$\\[3pt]
$\mu_\text{VBF+Vh}^{\gamma \gamma}$ & $1.05\pm 0.43$  & $-0.30$ & 1\\[3pt]
\hline
$\mu_\text{ggF+tth}^{bb}$ & $1.15\pm 0.97$  & $1$ & $4.5\cdot 10^{-3}$\\[3pt]
$\mu_\text{VBF+Vh}^{bb}$ & $0.65\pm 0.30$  & \:$4.5\cdot 10^{-3}$ & $1$\\[3pt]
\hline
$\mu_\text{ggF+tth}^{\tau \tau}$ & $1.06\pm 0.58$  & 1 & $-0.43$\\[3pt]
$\mu_\text{VBF+Vh}^{\tau \tau}$ & $1.12\pm 0.36$  & $-0.43$ & 1\\[3pt]
\hline
$\mu_\text{ggF+tth}^{WW}$ & $0.98\pm 0.21$  & 1 & $-0.14$\\[3pt]
$\mu_\text{VBF+Vh}^{WW}$ & $1.38\pm 0.39$  & $-0.14$ & 1\\[3pt]
\hline
$\mu_\text{ggF+tth}^{ZZ}$ & $1.42\pm 0.35$  & 1 & $-0.49$\\[3pt]
$\mu_\text{VBF+Vh}^{ZZ}$ & $0.47\pm 1.37$  & $-0.49$ & 1\\[3pt]
\hline
\end{tabular}
\caption{$h$ signal strengths from Fig.~13 and Table 14 of \cite{Khachatryan:2016vau}.}
\label{tab:signalstrengths}
\end{table}

\begin{table}
\begin{tabular}{| c | c | c | c |}
\hline
\textbf{Channel} & \textbf{Experiment} & \textbf{Source} &  \textbf{Mass range (GeV)}\\
\hline
\multirow{2}{*}{$gg\to H/A \to \tau\tau$} & ATLAS & Fig.~11a of \cite{Aad:2014vgg}  & 90-1000 \\
& CMS & Fig.~10 (left) of \cite{CMS:2015mca}  & 90-1000 \\
\hline
\multirow{2}{*}{$b\bar{b}\to H/A \to \tau\tau$} & ATLAS & Fig.~11b of \cite{Aad:2014vgg}  & 90-1000 \\
& CMS & Fig.~10 (right) of \cite{CMS:2015mca}  & 90-1000 \\
\hline
\multirow{2}{*}{$gg\to H/A \to \gamma\gamma$} & ATLAS & Fig.~4, \cite{Aad:2014ioa} & 65-600 \\
& CMS & Fig.~7 (left) of \cite{Khachatryan:2015qba} & 150-850 \\
\hline
$b\bar{b}\to H/A \to b\bar{b}$ & CMS & Fig.~6 of \cite{Khachatryan:2015tra} & 100-900 \\ 
\hline
$gg\to H\to WW$ & ATLAS &  Fig.~13 (left) of \cite{Aad:2015agg} & 300-1500
\\ \hline
$WW/ZZ \to H\to WW$ & ATLAS &  Fig.~13 (right) of \cite{Aad:2015agg} & 300-1500 \\
\hline
$gg\to H\to ZZ$ & ATLAS &  Fig.~12a of \cite{Aad:2015kna} & 140-1000 \\
\hline
$WW/ZZ \to H\to ZZ$ & ATLAS &  Fig.~12b of \cite{Aad:2015kna} & 140-1000 \\
\hline
$pp\to H\to ZZ$\footnote{Signal strength (normalized to the SM expectation)} & CMS & Fig.~7 (bottom right) of \cite{Khachatryan:2015cwa} & 150-1000 \\
\hline
$gg\to H\to hh$ & ATLAS & Fig.~6 of \cite{Aad:2015xja} & 260-1000 \\
\hline
$pp\to H\to hh [\to (b\bar{b}) (\tau\tau)]$ & CMS & Fig.~5a of \cite{CMS:2016zxv} & 300-1000 \\
\hline
$pp\to H\to hh \to (\gamma \gamma) (b \bar{b})$ & CMS & Fig.~8 of \cite{Khachatryan:2016sey} & 250-1100 \\
\hline
$pp\to H\to hh \to (b\bar{b}) (b\bar{b})$ & CMS &  Fig.~5 (left) of \cite{Khachatryan:2015yea} & 270-1100 \\
\hline
$gg\to A\to hZ \to (\tau\tau) (\ell \ell)$ & CMS & Fig.~10  (left) of \cite{Khachatryan:2015tha} & 220-350 \\
\hline
$gg\to A\to hZ \to (b\bar{b}) (\ell \ell)$ & CMS & Fig.~3 of \cite{Khachatryan:2015lba} & 225-600 \\
\hline
$gg\to A\to hZ\to (\tau\tau) Z$ & ATLAS & Fig.~3a of \cite{Aad:2015wra}  & 220-1000 \\
\hline
$gg\to A\to hZ\to (b\bar{b})Z$ & ATLAS & Fig.~3b of \cite{Aad:2015wra} & 220-1000 \\
\hline
$pp\to A\to Z\gamma \to (\ell \ell)\gamma$ & CMS & Fig.~2 of \cite{CMS:2016all} & 200-1200 \\
\hline
\end{tabular}
\caption{The exclusion (upper) limits at $ 95\% $ CL on the production cross-section times branching ratio of the processes considered in the $ H $ and $ A $ searches. The first four exclusion limits are employed in both, $ H $ and $ A $ searches.}
\label{tab:heavysearches}
\end{table}

\begin{table}
\begin{tabular}{| c | c | c |}
\hline
\textbf{Observable} & \textbf{Value} & Source\\
\hline
$\Delta m_{B_s}$ & $17.757\pm 0.021$ ps$^{-1}$ & \cite{Amhis:2014hma}\\
\hline
$\mathcal{B}(\bar{B}\to X_s\gamma)$ & $3.43\cdot 10^{-4} \pm 0.21\cdot 10^{-4} \pm 0.07\cdot 10^{-4}$  & \cite{Amhis:2014hma}\\
\hline
\end{tabular}
\caption{Flavour inputs.}
\label{tab:Flavour}
\end{table}

%
\bibliographystyle{utphys}
\bibliography{NLOuni}

\end{document}